\documentclass[pdftex,headsepline,twoside,11pt]{article}
\usepackage[T1]{fontenc} 
\usepackage{paralist,textcomp,helvet,graphicx,textfit,eurosym,url,longtable,float,pdflscape,comment,rotating,rotfloat,comment,bm,moreverb,amsfonts,listings}
\usepackage[centertags]{amsmath}
\usepackage{pgf}
\usepackage{array}
\usepackage[textfont=scriptsize]{subcaption}
\usepackage[absolute]{textpos}
\usepackage[paper=a4paper,left=25mm,right=25mm,top=25mm,bottom=25mm]{geometry}
\usepackage{cite}
\usepackage{blkarray}

\usepackage{pgf}
\usepackage{array}
\usepackage[textfont=scriptsize]{subcaption}
\usepackage[absolute]{textpos}
\usepackage{cite}
\usepackage{blkarray}
\usepackage{epstopdf}
\usepackage[nonumberlist]{glossaries}

\newcommand\BibTeX{{\rmfamily B\kern-.05em \textsc{i\kern-.025em b}\kern-.08em
T\kern-.1667em\lower.7ex\hbox{E}\kern-.125emX}}

\definecolor{mygreen}{rgb}{0.4,0.4,0.4}
\definecolor{mygray}{rgb}{0,0,0}
\definecolor{mymauve}{rgb}{0,0,0}

\makeglossaries
\newglossaryentry{ANOVA}
{
  name=ANOVA,
  description={Analysis of Variance}
}

\newglossaryentry{BCEA}
{
  name=BCEA,
  description={Bayesian Cost-Effectiveness Analysis}
}

\newglossaryentry{BMM}
{
  name=BMM,
  description={Bayesian Markov Model}
}

\newglossaryentry{BODE}
{
  name=BODE,
  description={Bayesian ODE-based model}
}

\newglossaryentry{CEA}
{
  name=CEA,
  description={Cost-Effectiveness Analysis}
}

\newglossaryentry{CEAC}
{
  name=CEAC,
  description={Cost-Effectiveness Acceptability Curve}
}

\newglossaryentry{CI}
{
  name=CI,
  description={Credible Interval}
}

\newglossaryentry{DDR}
{
  name=DDR,
  description={Double Data Rate}
}

\newglossaryentry{dODE}
{
  name=dODE,
  description={deterministic ODE-based model}
}

\newglossaryentry{EVPI}
{
  name=EVPI,
  description={Expected Value of Perfect Information}
}

\newglossaryentry{EVPPI}
{
  name=EVPPI,
  description={Expected Value of Partial Perfect Information}
}

\newglossaryentry{EVSI}
{
  name=EVSI,
  description={Expected Value of Sample Information}
}

\newglossaryentry{FAST}
{
  name=FAST,
  description={Fourier Amplitude Sensitivity Test}
}

\newglossaryentry{GB}
{
  name=GB,
  description={Gigabyte}
}

\newglossaryentry{HDD}
{
  name=HDD,
  description={Hard Disk Drive}
}

\newglossaryentry{HIV}
{
  name=HIV,
  description={Human Immunodeficiency Virus}
}

\newglossaryentry{HMC}
{
  name=HMC,
  description={Hamiltonian Monte Carlo}
}

\newglossaryentry{HPV}
{
  name=HPV,
  description={Human papillomavirus}
}

\newglossaryentry{ICER}
{
  name=ICER,
  description={Incremental Cost-Effectiveness Ratio}
}

\newglossaryentry{IPP}
{
  name=IPP,
  description={Independent Inspection Process}
}

\newglossaryentry{ISPOR}
{
  name=ISPOR,
  description={International Society for Pharmacoeconomics and Outcomes Research}
}

\newglossaryentry{JAGS}
{
  name=JAGS,
  description={Just Another Gibbs Sampler}
}

\newglossaryentry{MCMC}
{
  name=MCMC,
  description={Markov Chain Monte Carlo}
}

\newglossaryentry{MM}
{
  name=MM,
  description={Markov Model}
}

\newglossaryentry{NB}
{
  name=NB,
  description={Monetary Net Benefit}
}

\newglossaryentry{NICE}
{
  name=NICE,
  description={National Institute for Health and Care Excellence}
}

\newglossaryentry{ODE}
{
  name=ODE,
  description={Ordinary Differential Equation}
}

\newglossaryentry{PSA}
{
  name=PSA,
  description={Probabilistic Sensitivity Analysis}
}

\newglossaryentry{QALY}
{
  name=QALY,
  description={Quality-Adjusted Life Year}
}

\newglossaryentry{RAM}
{
  name=RAM,
  description={Random Access Memory}
}

\newglossaryentry{SATA}
{
  name=SATA,
  description={Serial Advanced Technology Attachment}
}

\newglossaryentry{STI}
{
  name=STI,
  description={Sexually Transmitted Infection}
}

\newglossaryentry{VoI}
{
  name=VoI,
  description={Value of Information}
}

\newglossaryentry{WBDiff}
{
  name=WBDiff,
  description={WinBUGS Differential Interface}
}

\newglossaryentry{WinBUGS}
{
  name=WinBUGS,
  description={Bayesian inference Using Gibbs Sampling (for Windows)}
}

\title{A dynamic Bayesian Markov model for health economic evaluations of interventions against infectious diseases}
\author{Katrin Haeussler, Ardo van den Hout, Gianluca Baio}

\begin{document}
\maketitle

\begin{abstract}

\textbf{Background} 
Health economic evaluations of interventions in infectious disease are commonly based on the predictions of ordinary differential equation (ODE) systems or Markov models (MMs). Standard MMs are static, whereas ODE systems are usually dynamic and account for herd immunity which is crucial to prevent overestimation of infection prevalence. Complex ODE systems including distributions on model parameters are computationally intensive. Thus, mainly ODE-based models including fixed parameter values are presented in the literature. These do not account for parameter uncertainty. As a consequence, probabilistic sensitivity analysis (PSA), a crucial component of health economic evaluations, cannot be conducted straightforwardly. 

\textbf{Methods} 
We present a dynamic MM under a Bayesian framework. We extend a static MM by incorporating the force of infection into the state allocation algorithm. The corresponding output is based on dynamic changes in prevalence and thus accounts for herd immunity. In contrast to deterministic ODE-based models, PSA can be conducted straightforwardly. We introduce a case study of a fictional sexually transmitted infection and compare our dynamic Bayesian MM to a deterministic and a Bayesian ODE system. The models are calibrated to simulated time series data.

\textbf{Results}
By means of the case study, we show that our methodology produces outcome which is comparable to the ``gold standard'' of the Bayesian ODE system.

\textbf{Conclusions}
In contrast to ODE systems in the literature, the dynamic MM includes distributions on all model parameters at manageable computational effort (including calibration). The run time of the Bayesian ODE system is 15 times longer.
\end{abstract}

\section{Background}

Vaccines, antibiotics and antivirals against infectious diseases offer health benefits to society \cite{Orenstein,Witty} and have been instrumental in the prevention and treatment of conditions previously causing egregious burden to public health. Examples include the extremely low prevalence of syphillis, the control of human immunodeficiency virus, the worldwide eradication of smallpox \cite{Weiss} and the extremely low incidence of tetanus, diphteria and congenital rubella syndrome in the Western world \cite{WHO3}. However, despite being frequently successful from a clinical point of view, vaccination programmes and antiretrovirals are often costly to apply. As a pre-requisite to their implementation, health interventions such as vaccines are thus increasingly subject to cost-effectiveness analyses (CEAs) \cite{WHO2,Briggs}.

The National Institute for Health and Care Excellence (NICE) is arguably the leading health technology assessment agency in the world. In the UK, NICE is responsible for providing guidance and advice on whether proposed interventions should be publicly funded. NICE has developed a set of criteria and guidelines that drive the analytic process of CEA \cite{NICE}. Crucially, these involve the explicit necessity of assessing the impact of parameter uncertainty on the decision making outcome, a process typically known as \textit{Probabilistic Sensitivity Analysis}~(PSA)~\cite{Briggs,Baio3,Baio2}. 

In the UK, the appraisal of vaccines falls under the remit of the Joint Committee for Vaccines and Immunisations, an independent expert advisory committee to the ministers and health departments. Since 2009, the Health Protection Regulation obliges the Secretary of State to ensure that recommendations for national vaccination programmes are based on an assessment demonstrating cost-effectiveness \cite{JCVI}. However, there are currently no vaccine-specific guidelines for developing clinical or cost-effectiveness evidence. 

One of the reasons for this circumstance is perhaps the intrinsic complexity of infectious disease modelling, which is typically performed through \textit{compartmental} models. These are highly complicated mathematical tools capable of simulating the natural history of disease infection and progression. More specifically, in pathogens transmissible among humans, these models need to account for population dynamics and \textit{herd immunity} \cite{Anderson3}. Herd immunity implies that due to lower infection prevalence, the introduction of preventive and therapeutic measures such as vaccination, quarantine, antivirals and antibiotics induces a reduced risk of pathogen exposure. Only dynamic models are able to prevent incorrect predictions since they are suitable to incorporate these effects \cite{Pitman,Edmunds2,Chong}. 

Dynamic compartmental models are commonly fitted by solving systems of Ordinary Differential Equations (ODEs) in continuous time. While these deterministic models usually deal with features such as herd immunity (and thus are considered the ``industry standard'' in infectious disease modelling), they are characterised by a notable computational effort. One important consequence is that, in most cases, epidemiological and economic modelling for infectious disease performed by means of ODEs is based on the inclusion of fixed, predefined values on the model parameters. These fixed values are usually informed through a point estimate. The joint uncertainty in the parameters is then not considered; these models result in outcome (e.g.\,on the number of people in the states) which does not include distributions. 

Therefore, PSA on the model outcome can only be conducted in retrospect and not in a straightforward way. An additional step using Latin Hypercube Sampling or Monte Carlo sampling is necessary, as shown in \cite{Blower,Jit,Jit2} and \cite{Khazeni,Alistar,Juusola}, respectively. Alternative methods that may prove computationally efficient when estimated through polynomial chaos expansions as shown in \cite{Xiu,Sudret} are provided by the Sobol and Fourier amplitude sensitivity test (FAST) indices. These indices are based on ANOVA techniques and thus estimate the total contribution of each model parameter or a combination of parameters to the variance of the model output \cite{Saltelli2,Saltelli}. However, in contrast to a full PSA, uncertainty is not propagated through the whole model.

The computational feasibility of PSA in retrospect is limited in models which include a high number of states and model parameters. In addition, in contrast to a Bayesian approach, parameter uncertainty is not propagated through the crucial model parts of pathogen transmission and disease progression. The outcome based on fixed parameter values can differ considerably from the PSA outcome, and the two are in most cases not reported in enough detail to identify possible inconsistencies (with the exception of \cite{Alistar}). The results presented commonly focus on uncertainty in the health economic rather than the prevalence outcome of the models, and PSA results on infection prevalence including the corresponding confidence intervals are often not given \cite{Khazeni,Zaric,Juusola}. This approach is highly questionable, especially with respect to consistency with validation targets. To ensure that the model outcome on the number of people in the states and infection prevalence is realistic, calibration to high quality data based on large sample size is necessary. This is often not conducted \cite{Zaric}, or only conducted on the outcome based on fixed parameter values \cite{Khazeni}; the PSA outcome on prevalence is usually not evaluated with respect to fitting high quality data \cite{Alistar,Juusola,Long}.                                                                                        

The more complex an ODE system especially with respect to state space and number of model parameters, the larger the effort on implementation and computation, especially if each model parameter is assigned a suitable distribution. This might be one of the reasons why the \textit{International Society for Pharmacoeconomics and Outcomes Research} guideline for best modelling practice in infectious disease suggests that PSA is \textit{not} a fundamental component of health economic assessment~\cite{Pitman}. This recommendation is given in contrast with NICE and virtually any other disease area. As a consequence, most economic models for vaccines only consider deterministic sensitivity analysis, which is based on selecting a grid of ``plausible'' values for a subset of model parameters in order to assess the robustness of the decision-making process. This approach is however not recommended in general, as it fails to account for potential correlation among the parameters \cite{Briggs,Andronis,Baio2}.

In contrast to ODE-based models, systems of equations can also be defined in discrete-time, which are termed difference equation models \cite{Vynnycky}. An alternative compartmental specification is given by Markov models (MMs). MMs are used to model progression over time across a finite set of states. Since MMs are described by a stochastic process, these are classified as stochastic models. This is in contrast to ODE systems and difference equation models, which belong to the class of deterministic models. In a deterministic model, the same set of parameter values and initial conditions always results in the same output. In contrast, a stochastic model produces different output each time the model is run, accounting for randomness. Apart from the model class, difference equation models and discrete-time MMs are mathematically comparable.

Although MMs can also be computationally intensive, it is generally feasible to implement even complex models in a Bayesian framework or to use re-sampling methods such as the bootstrap to characterise the uncertainty in the model parameters. Perhaps for this reason, MMs are a very popular tool in health economic evaluation. Nevertheless, a major limitation in infectious disease modelling is that they are intrinsically static, i.e.\ they do not account for \mbox{population dynamics} \cite{Brisson}.

We introduce in this paper an extension to standard MMs, which we term ``dynamic Bayesian MM'' to indicate that we consider a stochastic model and use a Bayesian framework to estimate its underlying parameters. We directly include the force of infection of the pathogen, which automatically accounts for time-dependent changes in prevalence and thus the effects of herd immunity, into the state allocation algorithm of a MM. In other words, the movement of susceptibles to the state of infection is directly represented by the dynamic force of infection. A direct inclusion of the force of infection into the state allocation algorithm of a difference equation model \cite{Ross}, a MM \cite{Cooper2,Gibson2} or its direct consideration in a model based on a stochastic process \cite{Forrester,Auranen2,Gibson} was presented previously by several authors. However, the authors who present stochastic models do not conduct a health economic evaluation. In the health economics literature, to the best of our knowledge, no approach of a dynamic MM including a high number of states and suitable probability distributions on all model parameters is presented. 

Our dynamic Bayesian MM combines six advantages in comparison to the few dynamic MMs presented in the literature. Firstly, in contrast to our contribution, the compartmental models in the literature are only suitable to include a low number of states due to computational limitations and the majority consist of no more than four states (apart from \cite{Ross}). Our methodology is especially suitable to incorporate an extensive number of states as described elsewhere \cite{Haeussler} for the application to human papillomavirus (HPV) modelling. In our model on HPV, 36 states in females and 22 states in males are included to account for all known HPV-induced diseases apart from recurrent respiratory papillomatosis. In addition, statisticians or health economic modellers can implement our approach directly in the commonly used software \texttt{R} linking to \texttt{JAGS} or \texttt{WinBUGS}; the corresponding run-time is considerably fast, and therefore, it is not necessary to use a compiled language such as \texttt{C} which is usually not widely used in this field. The four remaining advantages are given through the Bayesian framework, which \textit{i)} is highly flexible with its probabilistic nature since it considers multiple sources of prior information in terms of evidence synthesis \cite{Spiegelhalter}, \textit{ii)} enables propagation of parameter uncertainty through the infection transmission, progression and economic evaluation process, \textit{iii)} ensures that calibration targets are met through a constant updating process of the outcome on the numbers of people in the states directly in the state allocation algorithm, using available time series data, and \textit{iv)} simplifies the process of PSA, an essential part of CEAs, avoiding the necessity of applying additional sampling techniques such as Monte Carlo or Latin Hypercube Sampling once the model output is available. 

The paper is structured as follows: Firstly, we describe compartmental models in widespread use for CEAs on interventions against infectious diseases. Secondly, we introduce our contribution of the dynamic Bayesian MM. Thirdly, we compare the performance of an ODE system including fixed values on the model parameters and an ODE system in a Bayesian framework to our methodology, using a case study of a chronic sexually transmitted infection. To contrast the three methodologies in practice, we compare the natural history of disease following calibration, and conduct CEAs including PSA, comparing a screening strategy to a hypothetical vaccine. Finally, we discuss advantages and disadvantages of ODE-based methodology in comparison to the dynamic Bayesian MM.

\section{Methods\label{ode}} 

Compartmental models consist of a set $\mathcal{S}$ of mutually exclusive and exhaustive states describing disease infection and progression. We indicate the elements of $\mathcal{S}$ as $s=1,\ldots,S$. Members of a ``virtual'' population move across the states over a pre-specified time horizon. 

\begin{figure}[H]
\centering
\includegraphics[width=\textwidth]{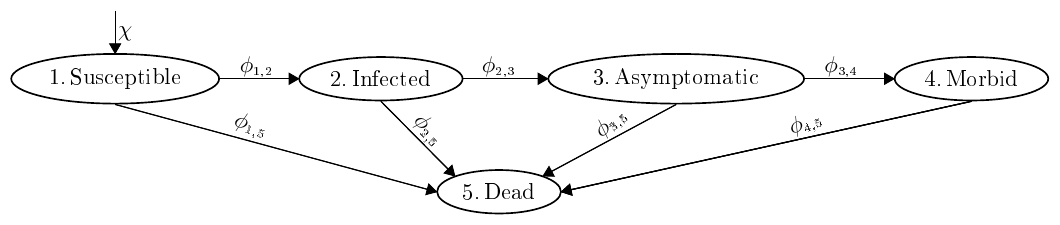}
\caption{Model structure of a hypothetical chronic sexually transmitted infection. The arrows represent the possible transitions. These are governed by the parameters $\phi_{r,s}$ with indices $r,s\in\mathcal{S}$ representing origin and target states, respectively. The replenishment of the pool of susceptibles by newborns proceeds at a rate~$\chi$.}\label{fig:SIODE}
\end{figure}

Figure \ref{fig:SIODE} shows an example of a compartmental model incorporating the natural disease history of a chronic sexually transmitted infection (STI) with $S=5$ states which is deemed similar to HIV \cite{Alizon}. The assumptions encoded by this structure are that the whole population initially is in the state \textit{Susceptible} (indexed by $s=1$), from which a proportion can move to the state \textit{Infected} ($s=2$). Following this, people move to an \textit{Asymptomatic} state ($s=3$). A progression to the state \textit{Morbid} ($s=4$) induces the development of disease symptoms. The state \textit{Dead} ($s=5$) can be reached from any state; people die due to any cause or as a consequence of being in the state \textit{Morbid}. Compared to the average population, the latter have a higher risk of death. A transition from one state to another is defined according to \textit{transition parameters} \cite{Welton}. They are indicated as $\phi_{r,s}$, where $r,s\in \mathcal{S}$ represent the original and target state, respectively. People proliferate at a rate $\chi$, resulting in a replenishment of the pool of susceptibles at risk of contracting the infection.

Compartmental models may differ in two characteristics. The first is the specification of time. The most realistic option is to allow transitions among the states at any point in time $t>0$; this is a so-called ``continuous-time approach''. Alternatively, it is possible to assume that transitions occur in discrete time where only one transition is possible within a pre-defined time interval $\mathcal{I}_t = [t, t + \kappa)$, where $\kappa$ determines the corresponding interval width, commonly referred to as \textit{cycle}. Depending on the medical context, $\kappa$ can be specified in terms of daily, weekly, monthly or yearly~cycles. The second difference concerns the way in which population dynamics are considered: models including a force of infection which accounts for changes in prevalence are referred to as \textit{dynamic}, while those that include a fixed force of infection and thus ignore the effects of herd immunity are termed \textit{static}. 

In addition, different approaches to model parameter specification exist, which may have major impact on the outcome of health economic evaluations. Depending on the methodology used, the induced computational effort might not allow the inclusion of probability distributions on all model parameters. In that case, the model parameters are fixed values. Commonly, these are estimated through a relevant summary, e.g. mean, median or mode, obtained from available data. The point estimate is then used as a plug-in for the corresponding parameter. In contrast, if whole distributions of values are assigned, parameter uncertainty is propagated through the infection~progression. While frequentist versions of this strategy exist (e.g.\ based on bootstrap), this type of modelling is most naturally handled within a Bayesian~paradigm.

\subsection{Ordinary Differential Equation models\label{odecase}} 

ODE systems model the rate of change in the number of people within a given state in continuous time; thus, the corresponding parameters are transition \textit{rates} and we denote them as $\rho_{r,s}(t)$, with $r,s\in\mathcal{S}$ representing again the origin and target states, respectively. In principle, transition rates can depend on $t$, but do not necessarily have to. The number of people transitioning in each state at $t$ is multiplied by the corresponding transition rates to obtain the inflow and outflow to and from a state. The difference between the number of people entering and leaving a state corresponds to the derivative of the number of those in the respective state. 

Back to our example, we define the vector $\bm{n}(t)=\left(n_1(t),\ldots,n_S(t)\right)'$, where $n_s(t)$ is the number of people in state $s$ at time $t$. The corresponding ODE system is given by the set of equations
\begin{equation}\label{ode_chronic}
\begin{split}
\frac{dn_1(t)}{dt}&=\chi[n_1(t)+n_2(t)+n_3(t)+n_4(t)]-\rho_{1,2}(t)n_1(t)-\rho_{1,5} n_1(t)\\
\frac{dn_2(t)}{dt}&=\rho_{1,2}(t)n_1(t)-\rho_{2,3} n_2(t)-\rho_{2,5} n_2(t)\\
\frac{dn_3(t)}{dt}&=\rho_{2,3} n_2(t)-\rho_{3,4} n_3(t)-\rho_{3,5} n_3(t)\\
\frac{dn_4(t)}{dt}&=\rho_{3,4} n_3(t)-\rho_{4,5} n_4(t)\\
\frac{dn_5(t)}{dt}&=\rho_{1,5} n_1(t)+\rho_{2,5} n_2(t)+\rho_{3,5} n_3(t)+\rho_{4,5} n_4(t).
\end{split}
\end{equation}

\noindent
The rate of change in the number of people in each state at each point in time $t$ is subject to population dynamics and exposure to sources of infection. The transition to the state of infection is determined by the dynamic force of infection of the pathogen, indicated by $\rho_{1,2}(t)$ in~(\ref{ode_chronic}). This is a function of the probability of pathogen transmission, partner acquisition rates and prevalence.  

If each model parameter is assigned a fixed value, parameter uncertainty is not accounted for. Scenario analyses are often performed, for example by estimating the parameters through summary statistics at the extremes of the corresponding parameter distribution (e.g. lower or upper quantiles, minima or maxima). As mentioned earlier, this is not equivalent to the application of a full PSA.

Theoretically, it is possible to incorporate a probability distribution for each parameter of an ODE-based model, for example in a Bayesian context. The uncertainty is then propagated through the estimation procedure, which again generates a full distribution of outcomes. This type of model can be analysed using for instance software based on Markov Chain Monte Carlo (MCMC) algorithms such as \texttt{WinBUGS} \cite{WinBUGS} 
or \texttt{Stan} \cite{stan}, a very promising tool, which in general performs extremely well with relatively complex systems. Both include ODE solvers and can be linked to the statistical programming language~\texttt{R}.

In realistic problems including a large number of states and complex structures, assigning a suitable distribution to all model parameters in ODE systems may be impractical since the model needs to be run for a large number of simulations to ensure convergence of each parameter and thus the ODEs have to be solved repeatedly for each parameter combination. The increase in the computational time is induced by the length of the observation time horizon, the amount of parameters, the complexity of contacts, and most importantly by the number of states which increases the number of differential equations. Consequently, complex ODE-based models which incorporate distributions on all parameters are rare exceptions in the literature on infectious disease transmission modelling \cite{Bilcke}.

\subsection{Discrete-time Markov models\label{movementsMM}}

The main characteristic of MMs is the Markov assumption which implies that the transition to a future state is exclusively conditional on the current state or on a limited set of previous states, but not on the full trajectory. However, the Markov assumption can be relaxed by accounting for covariates (e.g.\,age and sex) or for time-specific prevalence and population dynamics in the transition parameters. 

ODE systems and MMs differ in the way they describe the process of transitions. As suggested earlier, in the former, the rates of change are calculated dynamically through differentiation, while in the latter, the transitions are described by a static Markov process (a category of a stochastic process). As for ODE-based models, MMs can be implemented for continuous time. 

However, the vast majority of MMs in the health economic literature is based on a discrete-time approach \cite{Rosmalen}. In this case, members of the population move across the states according to a set of transition \textit{probabilities} only once per time interval (termed ``Markov cycle''). These probabilities can be arranged in a matrix $\bm{\Pi}=\left(\pi_{r,s}\right)$, whose elements represent the transition probabilities for movements from an original state $r$ to a target state $s$.

For the model structure of Figure \ref{fig:SIODE}, the transition probability matrix is defined as

\begin{equation*}\label{transprobmatrix}
\bm\Pi = \left(
\begin{array}{ccccc}
\pi_{1,1}& \pi_{1,2} & 0 & 0& \pi_{1,5}\\
0& \pi_{2,2}& \pi_{2,3} & 0& \pi_{2,5}\\
0& 0 & \pi_{3,3} & \pi_{3,4}& \pi_{3,5}\\
0& 0 & 0 & \pi_{4,4} & \pi_{4,5}\\
0& 0 & 0 & 0& 1\\
\end{array}
\right),
\end{equation*}

\noindent
implying that, for example, a susceptible either acquires the infection (with probability $\pi_{1,2}$), dies (with probability $\pi_{1,5}$), or remains susceptible, which occurs with probability $\pi_{1,1} = 1-\pi_{1,2}-\pi_{1,5}$. 

If we define the vector $\bm{n}_t=\left(n_{1t},\ldots,n_{St}\right)'$, where $n_{st}$ is the number of people in state $s$ and at each time interval~$\mathcal{I}_t$, then transitions across the states from one time interval to the next are calculated as 

\begin{equation}\label{allocation}
\bm{n}_{t+1}=\bm\Pi\bm{n}_t. 
\end{equation}

MMs are relatively straightforward to implement and are commonly used to model the progression of non-communicable conditions such as cardiovascular disease and cancer. Therefore, they are established in the health economic literature and well-known to clinicians and decision makers. However, the process of pathogen transmission is not estimated correctly using standard MMs. A transition of susceptibles to the state of infection is commonly represented by a static transition parameter which does not consider changes in the population prevalence over time. These especially occur after the introduction of a preventive intervention such as vaccination into a fully susceptible population. 

The predictions of static MMs on population prevalence are commonly incorrect (although notable exceptions include scenarios with very low vaccine coverage or pathogens that cannot be transmitted between humans, e.g.\ tetanus). In the worst case, the whole model outcome on infection prevalence and the related CEA can be incorrect, e.g.\ because of the impact of an unrecognised shift in the age of infection of childhood diseases. Some childhood diseases are relatively harmless in young children but prone to lead to serious health issues in adults. Incorrect predictions of static MMs on population health and induced costs, e.g.\ through hospitalisation and treatment, can have dire consequences \cite{Brisson}. 

As for ODE-based models, a dynamic force of infection could be incorporated into the transition probabilities to account for the effects of herd immunity. However, to the best of our knowledge, dynamic Markov models are not commonly used in the health economics literature.

\subsection{Dynamic Bayesian Markov models\label{methodology}}

To overcome the limitations discussed above and with a view to extending the modelling framework for health economic evaluation of interventions in infectious disease, our main idea is to add a force of infection which depends on population dynamics and prevalence into a MM setting. As a consequence, the transition probabilities from the state \textit{Susceptible} to the state \textit{Infected} are directly defined by the dynamic force of infection. Specifically, we set up our model so that the force of infection is calculated separately within each cycle of the state allocation algorithm corresponding to (\ref{allocation}) as a function~of 

\begin{itemize}
\item the probability of pathogen transmission per contact, which we indicate as $\beta$; 
\item the rate of contacts between susceptible and infectious members of the population $\omega$; and
\item the time-dependent pathogen prevalence \[ \psi_t = \frac{I_t}{N_t}, \] where $I_t$ represents the number of people in the state of infection and, assuming that state $S$ indicates death, \[N_t=\sum_{s=1}^{S-1}n_{st}\] is the number of those alive at time interval $\mathcal{I}_t$. 
\end{itemize} 

The force of infection is recalculated at each Markov cycle as 

\begin{equation}\label{forceinf2}
\lambda_t=\beta \omega \psi_t.
\end{equation}
Since $\omega$ is a rate, (\ref{forceinf2}) also results in a transition rate. Assuming that $\lambda_t$ remains constant within each time interval, the corresponding time-dependent transition probability for the discrete-time MM is estimated as

\begin{equation}
\pi_{\mbox{\tiny{1,2,t}}} = 1-\exp^{-\lambda_t}.\label{coop}
\end{equation}

We acknowledge that the estimation of the force of infection may be only approximate, due to the competing risk of death and the assumption of uniformity within the intervals $\mathcal{I}_t$. This assumption is not likely to hold if the disease is characterised by very fast transmission, or when events associated with the infection are likely to occur in short periods of time. In these cases, it is perhaps advisable to reduce the length $\kappa$ of the cycles and the duration of the follow-up, to avoid unrealistic estimates for the number of subjects in the states per cycle. For example, for yearly cycle length, immediate death following infection would be delayed by up to one year. These delays would then accumulate through the whole model and introduce a more substantial bias at later follow-up. 

However, a likelihood function for interval censored data could be estimated to account for competing risks as shown in \cite{Hudgens,Andersen} for an independent inspection process (IPP) model; this model allows for future movements to be conditional on the history of the data. In Lemma 1, Hudgens et al. \cite{Hudgens} present this likelihood under the assumption that only one event can occur per cycle. As an alternative to~(\ref{coop}), this likelihood could be derived for the cumulative distribution function of an exponential model. 

Moreover, competing risks could be considered through the Kolmogorov forward equations. This alternative approach would allow for the possibility that more than one event could occur per cycle; for instance, one could acquire the infection followed by death within the same cycle. In order to move from the state \textit{Susceptible} to \textit{Infected}, one would have to take into account that the infection was acquired and death did not occur within the remainder of that cycle. For the exponential distribution, the corresponding Kolmogorov forward equation to estimate the transition probability of moving between the two states was derived as 

\begin{equation}\label{kolg}
\pi_{\mbox{\tiny{1,2,t}}}=\exp^{-\rho_{\mbox{\tiny{1,5}(t)}}}\left(1-\exp^{-\lambda_t}\right),
\end{equation}

\noindent accounting for $\rho_{\mbox{\tiny{1,5}(t)}}$, the transition rate of moving from the state \textit{Susceptible} to \textit{Death}. Replacing (\ref{coop}) by (\ref{kolg}), the accuracy of the approximation of $\pi_{\mbox{\tiny{1,2,t}}}$ could improve. Notice, however, that in the case study presented in Section \ref{healthtrans}, the probability of death is very low in susceptibles (who are by definition in good health and of young age) and therefore this additional complication in the estimation of the transition probability is not necessary. The results of the corresponding discrete- and continuous-time models are comparable; the approximation in (\ref{coop}) is therefore deemed sufficient.

The transition probability $\pi_{\mbox{\tiny{1,2,t}}}$ approximated by (\ref{coop}) is multiplied by the proportion of the population in the state \textit{Susceptible} to provide an estimation of the contingent of movements to the state \textit{Infected}, effectively including dynamic, time-dependent changes in prevalence in the corresponding transitions.

The computational effort is reduced by fitting models that do not involve complex ODEs, while still allowing for mixing patterns within the population. Another potential advantage of the dynamic MM framework is that it is fairly simple to incorporate suitable probability distributions on all model parameters, even if the model is complex with an extremely large number of parameters and states. In contrast, the related computational effort in a comparable ODE-based model would be extremely high due to i) numerical integration, e.g. through the Runge-Kutta solver \cite{Dreyer}; ii) considerably smaller step sizes of ODE solvers when compared to the cycle length of discrete-time models; and iii) accounting for competing risks in the transition rates as elaborated above. The model can be easily extended to include a high number of age cohorts for infectious diseases with age-specific prevalence such as HPV; we present this application including 24 age cohorts at manageable computational effort elsewhere \cite{Haeussler}.

Accounting for parameter uncertainty is particularly relevant because, for obvious ethical and practical reasons, it is invariably difficult (if possible at all) to obtain and use experimental evidence to inform the pathogen transmission probability $\beta$ and the active contact rate $\omega$ --- arguably the crucial parameters. Often observational studies or expert opinions are the only available information with the consequence that large uncertainty remains over the most likely range, let alone the ``true'' value of the parameters. A Bayesian approach may provide great benefit in allowing this uncertainty to be fully propagated and perhaps in integrating different sources of evidence (e.g.\ using evidence synthesis \cite{Welton}); this indeed has been advocated for MMs in the health economics literature \cite{Welton,Cooperetal,Baio2}. In \cite{Korostil}, $\beta$ is assigned a Uniform distribution in the interval [0;1], which essentially amounts to allowing any value (between 0 and 100\%) as equally possible. We have performed extensive sensitivity analysis to this parameter in the HPV model and found that the prevalence outcome was highly sensitive to this parameter (results \mbox{not published).} 

In a Bayesian dynamic MM setting, it is possible to assign prior distributions to the parameters $(\beta,\omega)$ to represent the state of science --- if data are available, these are updated into posterior distributions although it is possible to still propagate uncertainty in the priors even when no data on pathogen transmission or active contacts  are observed. In addition, the quantity $\psi_t$ is estimated for each cycle as a function of transition probabilities, which can also be modelled using suitable distributions. This modelling process induces a probability distribution on $\psi_t$ and \textit{a fortiori} also on $\lambda_t$, which is defined as a function of the three random parameters $(\beta,\omega,\psi_t)$. Thus, the corresponding transition probabilities $\pi_{\mbox{\tiny{1,2,t}}}$ are modelled probabilistically, meaning that uncertainty in the population dynamics is propagated through the economic~model.

If $\beta$ and $\omega$ are varied simultaneously and difficult to inform directly using empirical evidence, the force of infection may be subject to issues of non-identifiability, i.e. it is possible that it cannot be distinguished to what extend differences in the corresponding outcome on the number of subjects in the state \textit{Infected} are induced by which parameter. Individuals become infected as a consequence of meeting an infected subject (determined by $\beta$) once or several times (determined by $\omega$). However, we note here that we are mainly interested in the overall number of subjects in the state \textit{Infected}, irrespective of the number of contacts that are necessary to result in an infection. Therefore, in the economy of our modelling framework, it is less important if changes in the number of subjects in the state \textit{Infected} are induced by $\beta$ or $\omega$. Nevertheless, we do acknowledge this issue and suggest careful consideration of the prior assumptions when investigating sensitivity of the results.

Another crucial aspect in infectious disease modelling (and more generally in statistical analysis) is that of calibration of the model output \cite{Gelman,Vanni}. The Bayesian framework enables the calibration of the numbers of people in the states directly in the state allocation algorithm, using available time series data for a specific time frame of follow-up. The corresponding details are explained in Section~\ref{accparun} and Appendix~\ref{cal}.

The BMM is generalizeable to any infectious disease which is transmitted between humans and where interventions (e.g.\,quarantine, vaccination, antibiotic or antiviral treatment) are available to reduce infection prevalence. It is especially suitable to include an extremely high number of states, model parameters with suitable distributions, and age cohorts as shown in \cite{Haeussler}.

To estimate the efficacy and cost-effectiveness of an intervention, only dynamic models predict realistic outcome and account for herd immunity in case of vaccination. Yet, any kind of intervention which reduces prevalence has a protective impact on susceptibles who do not directly receive it; this can be considered in the BMM.

Finally and specifically for the purpose of economic evaluation, the dynamic BMM has the advantage that PSA can be performed ``for free'', once the model output is produced. In a Bayesian framework, the MCMC simulations for all the model parameters can be combined to obtain a full characterisation of the uncertainty in the decision-making process. This can be post-processed (e.g.\ using the \texttt{R} package \texttt{BCEA} \cite{BCEA}) to produce relevant summaries such as the cost-effectiveness plane, the cost-effectiveness acceptability curve and the analysis of the value of information (see Section~\ref{cea}).

Our approach to a dynamic BMM provides PSA samples directly as part of the probabilistic model output. Therefore, in addition to simpler probabilistic sensitivity analysis (e.g. based on CEACs), the BMM enables the conduct of extensive Value of Information (VoI) analysis, which can be used to prioritise further research on key model inputs, currently driving uncertainty in the decision making process. For example, recent methods have been developed for fast computation of both the Expected Value of Partial Perfect Information (EVPPI) and the Expected Value of Sample Information (EVSI) \cite{Strong,Heath,Heath2} that use simulations from the model parameters obtained directly during the process of PSA. The BMM conforms with this structure and would thus be suitable for VoI analysis based on these methods, without the need to divorce the transmission model from the economic evaluation, as often happens in current practice.

\subsection{Case study\label{healthtrans}}

We consider again the fictional chronic STI described above and compare the dynamic Bayesian MM to both a deterministic and a Bayesian ODE system. We denote the three models as BMM, dODE and BODE, respectively. We evaluate whether our BMM produces results that are in line with the ``gold standard'' of the BODE. In the three models, we distinguish between sexes as well as high- and low-risk sexual behaviour. The duration of the follow-up is set at 100 years, with a yearly Markov cycle length. We consider a population size of 1,000,000 and initially assume that 600 people are infected, whereas the remainder are susceptible. Males amount to 50\% and the high-risk group to 20\% of the population; the sex ratio in the two risk groups is constant. The proportion of infected people in both sexes and risk groups is identical. We account for sex-specific differences in sexual behaviour, assuming higher partner acquisition rates in males. The population size changes due to births and deaths. We conduct our analysis for two competing health-care interventions. We assume that in the \textit{status-quo}, screening takes place at intervals of five years at a pre-defined rate to enable an early detection of the STI. For simplicity, we assume that under the \textit{vaccination} scenario no screening takes place. We assume that the vaccine is only effective before initial STI infection; thus, susceptibles are vaccinated at a specified vaccine uptake rate at intervals of five years. Following STI diagnosis, treatment is provided in both~interventions.

In the dODE and BODE, the force of infection (transition rate from the state \textit{Susceptible} to the state \textit{Infected}) shown in (\ref{forceinf2}) has to be adjusted to account for the covariates sex and behaviour with respect to infection prevalence. In the BMM, the sex- and behavioural-specific transition probability is estimated by transforming this adjusted transition rate according to (\ref{coop}). For simplicity, we exclusively account for random mixing. Further details are given in Appendix~\ref{appforce}.

\subsubsection{Model parameters and related distributions\label{accparun}}
In addition to the probability of STI transmission $\beta$ and the partner acquisition rates $\omega_{vb}$, the model contains a variety of parameters such as those determining the screening and vaccine coverage, the unit costs of STI diagnostics and treatment and the health utilities, which are relevant in context of the cost-effectiveness analysis. We assign informative priors to transition parameters and costs (defined in monetary units of \textsterling) and utilities. We specify the distributional assumptions so that the outputs of the health economic evaluation are within reasonable~ranges. We assign informative or minimally-informative priors to the remaining parameters and update these through simulated individual-level and aggregate data into the corresponding posteriors. Using simulated data is reasonable since the case study is fictional. For example, we pretend that data on partner acquisition rates $\omega_{vb}$ are available from a large data registry. To infer $\omega_{vb}$, we update informative Gamma priors into the corresponding posteriors through Poisson-Gamma models. Beta-Binomial models are used to infer probabilities (e.g. the vaccine coverage \,$\alpha$ and efficacy \,$\gamma$, which are informed by data). We assume to have access to vaccination data of 500 individuals of whom 450 have received the vaccine. In addition, we assume to have the information that in 450 out of 500 people who received the vaccine, it was effective.

Table \ref{tab:param} shows an overview of the model parameters $\bm{\theta}=\{\omega_{vb},\chi,\beta,\pi_{r,s},\tau,\xi,\gamma,\sigma,\bm{c},u_s\}$. The means and 95\% credible intervals are displayed for the BMM (the values in the BODE are comparable) and rounded to two decimal places. However, the 95\% CI of the parameter $\chi$ is in fact defined as [0.009420;0.010596]. Since the BMM is based on a discrete-time approach, the corresponding transition probabilities are modelled using Beta distributions. In contrast, the transition rates of the continuous-time BODE are modelled using Gamma distributions. The transition probabilities for movements from the states \textit{Susceptible}, \textit{Infected} and \textit{Asymptomatic} to \textit{Dead} are assumed as identical; thus, only $\pi_{\mbox{\tiny{1,5}}}$ is shown. 

Depending on the medical context, correlations between posterior distributions are possible and informative priors can result in quite different posterior distributions, if they are influenced by the updating process of other priors. This might occur in parameters which are not independent, for example transition parameters to certain states calculated by means of hierarchical models. In this respect, the transition probability to the state \textit{Dead} might be dependent on one to a less severe state such as \textit{Morbid}. Potential correlation is automatically accounted for in the PSA.

\begin{table}[H]
\fontsize{6.5}{8.5}\selectfont
\caption{Overview of the informative priors and the models used for updating informative and minimally-informative priors. The values are fictional and were chosen so as to produce most realistic prevalence outcome and cost-effectiveness results. \label{tab:param}}
\centering
\setlength{\tabcolsep}{1.5mm}
\begin{tabular}{lp{3cm}llrr}
\hline
\textbf{Parameter}&\textbf{Description}&\textbf{Distribution/model BMM}&\textbf{Distribution/model BODE}&\textbf{Mean}&\textbf{2.5/97.5\% percentiles}\\
\hline
$\omega_{M\!H}$&Partner acquisition rate (high-risk males)& Poisson-Gamma model &equivalent to BMM &9.10&[8.77;9.29] \\
$\omega_{M\!L}$&Partner acquisition rate (low-risk males)& Poisson-Gamma model&equivalent to BMM &2.98& [2.82;3.12] \\
$\omega_{F\!H}$&Partner acquisition rate (high-risk females)& Poisson-Gamma model&equivalent to BMM &9.00&[8.71;9.26] \\
$\omega_{F\!L}$&Partner acquisition rate (low-risk females)& Poisson-Gamma model&equivalent to BMM &1.96& [1.86;2.09] \\
$\chi$&Proliferation parameter & Gamma(1111.1,111111.1)& Gamma(1111.1,111111.1)&0.01& [0.01;0.01]\\
$\beta$&STI transmission probability per partnership& Beta-Binomial model&equivalent to BMM&0.16& [0.15;0.16]\\
$\pi_{\mbox{\tiny{2,3}}}$&Transition parameter from state $2$ to state $3$& Beta(5119.2, 1279.8)& Gamma(25600,32000)&0.80& [0.79;0.81]\\
$\pi_{\mbox{\tiny{3,4}}}$&Transition parameter from state $3$ to state $4$& Beta(1842.66, 18631.34)& Gamma(2025,22500)&0.09& [0.09;0.09]\\
$\pi_{\mbox{\tiny{4,5}}}$&Transition parameter from state $4$ to state $5$& Beta(1535.96, 36863.04)& Gamma(1600,40000)&0.04& [0.04;0.04]\\
$\pi_{\mbox{\tiny{1,5}}}$&Transition parameter from state $1$ to state $5$& Beta(156.171, 312186.6)& Gamma(156.25,312500)&\textless0.01& [\textless0.01;\textless0.01]\\
$\eta$&Probability of STI diagnosis& Beta-Binomial model&equivalent to BMM&0.90& [0.88;0.92]\\
$\sigma$&Screening probability & Beta-Binomial model&equivalent to BMM&0.90& [0.87;0.92]\\
$\alpha$&Vaccine coverage parameter& Beta-Binomial model&equivalent to BMM&0.90& [0.87;0.92]\\
$\gamma$&Vaccine efficacy parameter& Beta-Binomial model&equivalent to BMM&0.90& [0.87;0.92]\\
$c_{screen}$&Unit cost of screening in~\textsterling&Lognormal(2.996, 0.693)& equivalent to BMM& 25.39& [5.19;77.53]\\
$c_{vac}$&Unit cost of vaccination in~\textsterling& Lognormal(5.011, 0.01)& equivalent to BMM& 150.02& [147.14;152.98]\\
$c_{test}$&Unit cost of STI test in~\textsterling& Lognormal(2.996, 0.03)& equivalent to BMM& 20.01& [18.83;21.19]\\
$c_{blood}$&Unit cost of blood test in \textsterling& Lognormal(3.401, 0.03)& equivalent to BMM& 30& [28.26;31.79]\\
$c_{treat}$&Unit cost of treatment in~\textsterling& Lognormal(8.517, 0.015)& equivalent to BMM& 4999.78& [4853.56;5149.24]\\
$c_{dis}$&Unit cost of disease treatment in~\textsterling& Lognormal(9.210, 0.01)& equivalent to BMM& 9999.95& [9802.97;10198.10]\\
$c_{gp}$&Unit cost of visit to general practitioner in~\textsterling& Lognormal(3.912, 0.02)& equivalent to BMM& 50.01& [48.08;52.01]\\
$u_2$&Health utility of infected (min=0, max=1)& Beta(1469.3, 629.7)& equivalent to BMM& 0.70& [0.68;0.72]\\
$u_3$&Health utility of asymptomatic (min=0, max=1)& Beta(1439.4, 959.6)& equivalent to BMM& 0.60& [0.58;0.62]\\
$u_4$&Health utility of morbid (min=0, max=1)& Beta(629.7, 1469.3)& equivalent to BMM& 0.30& [0.28;0.32]\\
\hline
\end{tabular}
\end{table}

The dODE is calibrated through a frequentist probabilistic calibration approach \cite{Vanni,vandeVelde3}, whereas the BODE and BMM are calibrated through Bayesian calibration approaches \cite{deAngelis,Welton2}. The first involves the calculation of goodness-of-fit statistics. The advantage of the latter is that the model parameters can be inferred by fitting the models to data directly (in one step). Simulated time series data on the number of high-risk people in the states \textit{Susceptible}, \textit{Infected}, \textit{Asymptomatic} and \textit{Morbid} in the first five years of follow-up are used for calibration. A short observation time horizon of only a couple of years (five years in this case) is common in available time series data. These data are simulated by running the dODE under the \textit{status quo} for a follow-up period of five years; the input parameter values are informed through the means of the parameters of the~BODE. In many infectious diseases, only data on the number of infected individuals are available. The model code can easily be adapted to only calibrate the outcome on one of the states since every state is calibrated separately. Further details are given in Appendix~\ref{cal}.

The dODE is estimated using a combination of the \texttt{R} packages \texttt{EpiModel} \cite{EpiModel} and \texttt{deSolve} \cite{deSolve}. As for the BMM and BODE, we estimate the model parameters using a MCMC procedure; we run two chains with a total of 1,000 simulations after convergence. In this setting, convergence is sufficiently achieved with a Potential Scale Reduction $\hat{R}<1.1$ in all model parameters \cite{Gelman}, and there are no issues with autocorrelation. We fit the BODE in \texttt{WinBUGS} through the ODE solver interface \texttt{WBDiff} \cite{WBDiff}. The BMM is estimated using \texttt{WinBUGS} and \texttt{JAGS} \cite{jags}, an alternative, established software to perform Gibbs sampling, in order to compare computational efficiency.

The models are run on a Dell Latitude E6320 (Intel Core i5-2520M, 2x4GB DDR3 RAM, 500GB SATA HDD (2.5", 7200rpm)). The computation times are 4,480.19 seconds (around 1 hour 15 minutes) and 6,587.67~seconds (around 1 hour 50 minutes) for the dODE and BODE, respectively. Interestingly, the BMM runs much faster in \texttt{JAGS} (149.81~seconds) than in \texttt{WinBUGS} (449.42~seconds). This difference is perhaps due to the way in which the two programmes handle logical nodes, which are instrumental to defining the state allocation algorithm (see (\ref{allocation})). All run times include model calibration which considerably increases the computational effort in the ODE systems, but not for the BMM. The relevant model codes are presented in Appendix~\ref{Rcode}.

As discussed in Section \ref{methodology}, identifiability could also become an issue when the posteriors of $\beta$ and $\omega$ are multimodal, which may happen if certain parameter combinations are more likely to favour the data on partner acquisition rates and STI transmission probabilities. As a consequence, the sampling algorithm could get stuck in one of the local modes, resulting in biased summary statistics. We generate different initial values for each Markov chain to reduce the risk that parts of the posterior distribution are not visited when producing the relevant samples. In addition, we evaluate whether the two chains show proper mixing through trace plots.

\section{Results}
\subsection{Natural history of disease\label{nathiscal}}

\begin{figure}[H]
\begin{subfigure}{.5\textwidth}
\centering
\includegraphics[width=0.9\linewidth]{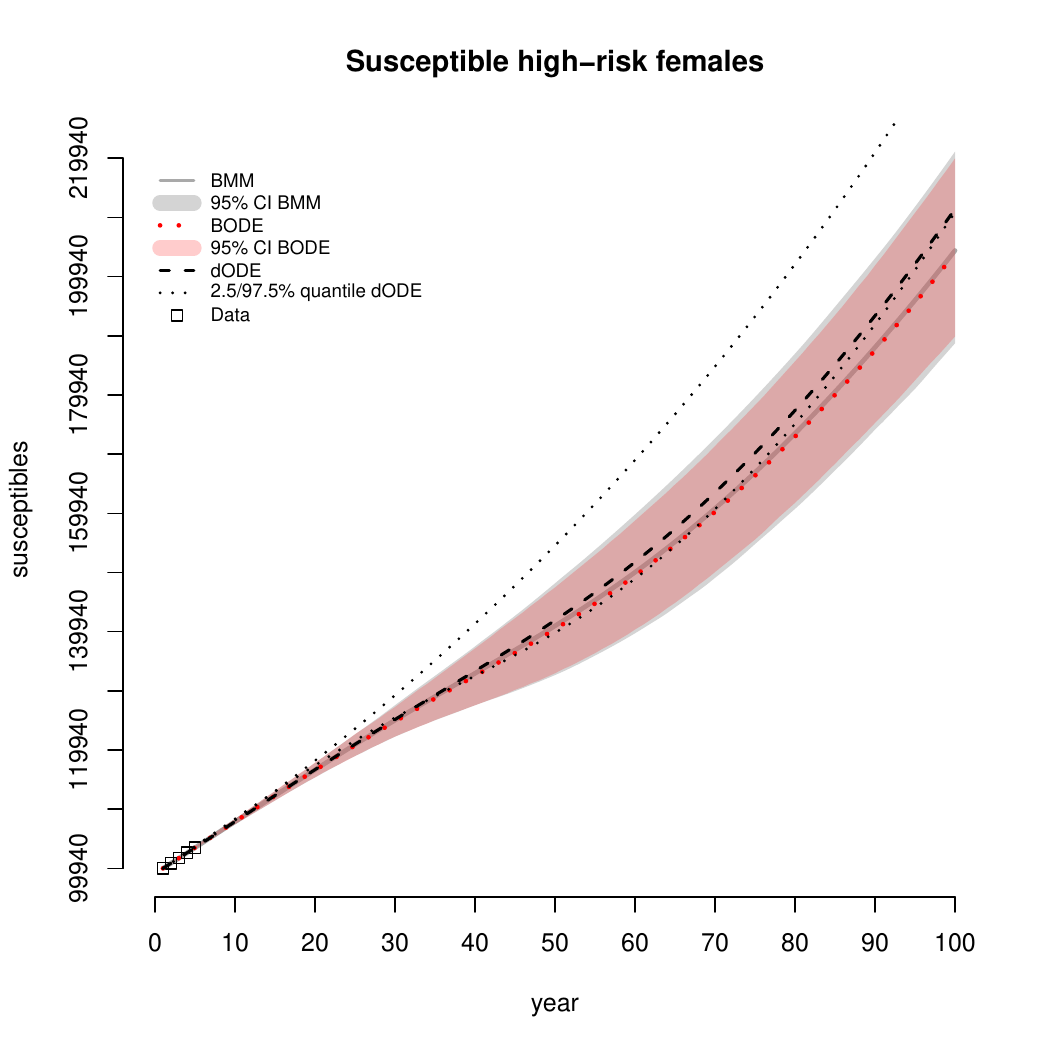}
\end{subfigure}
\hspace{0.1cm}
\begin{subfigure}{.5\textwidth}
\centering
\includegraphics[width=0.9\linewidth]{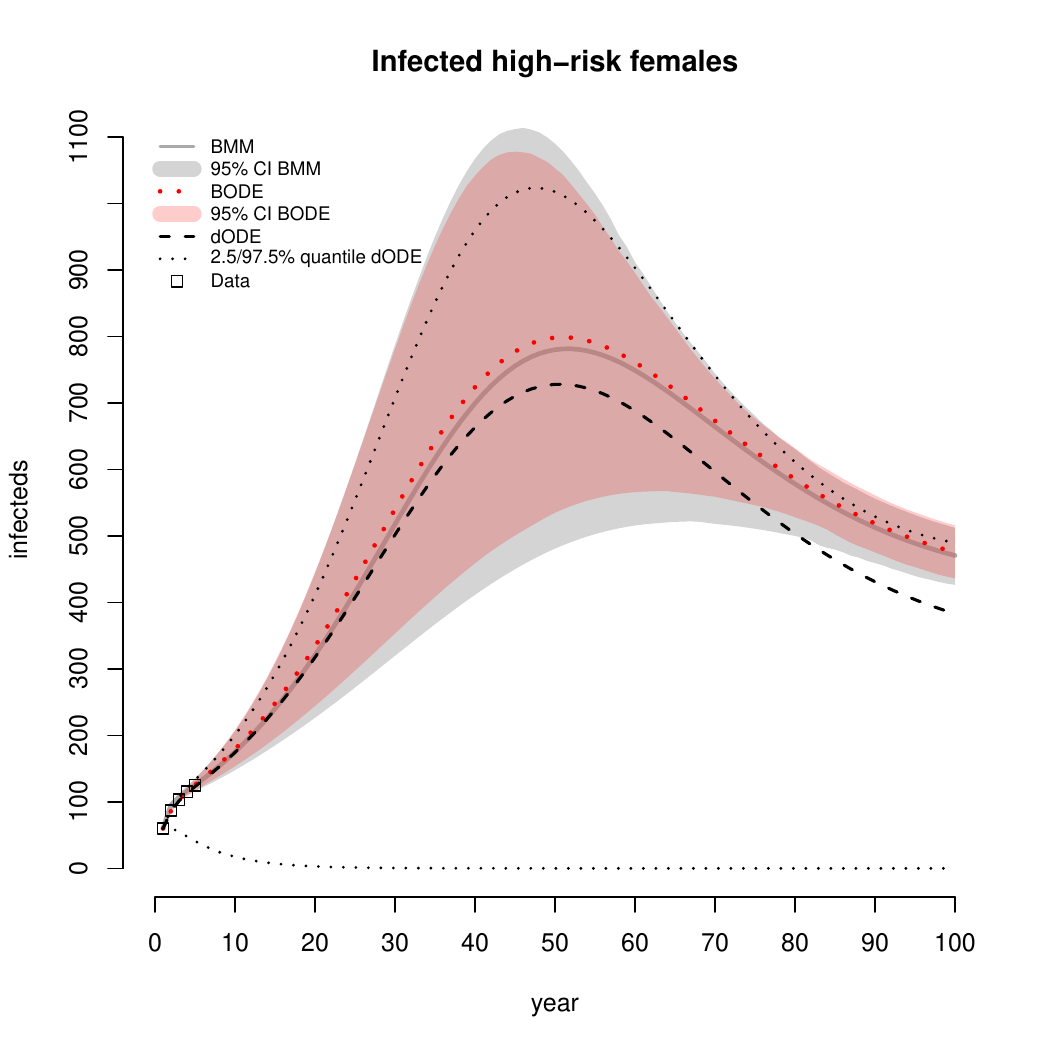}
\end{subfigure}
\begin{subfigure}{.5\textwidth}
\centering
\includegraphics[width=0.9\linewidth]{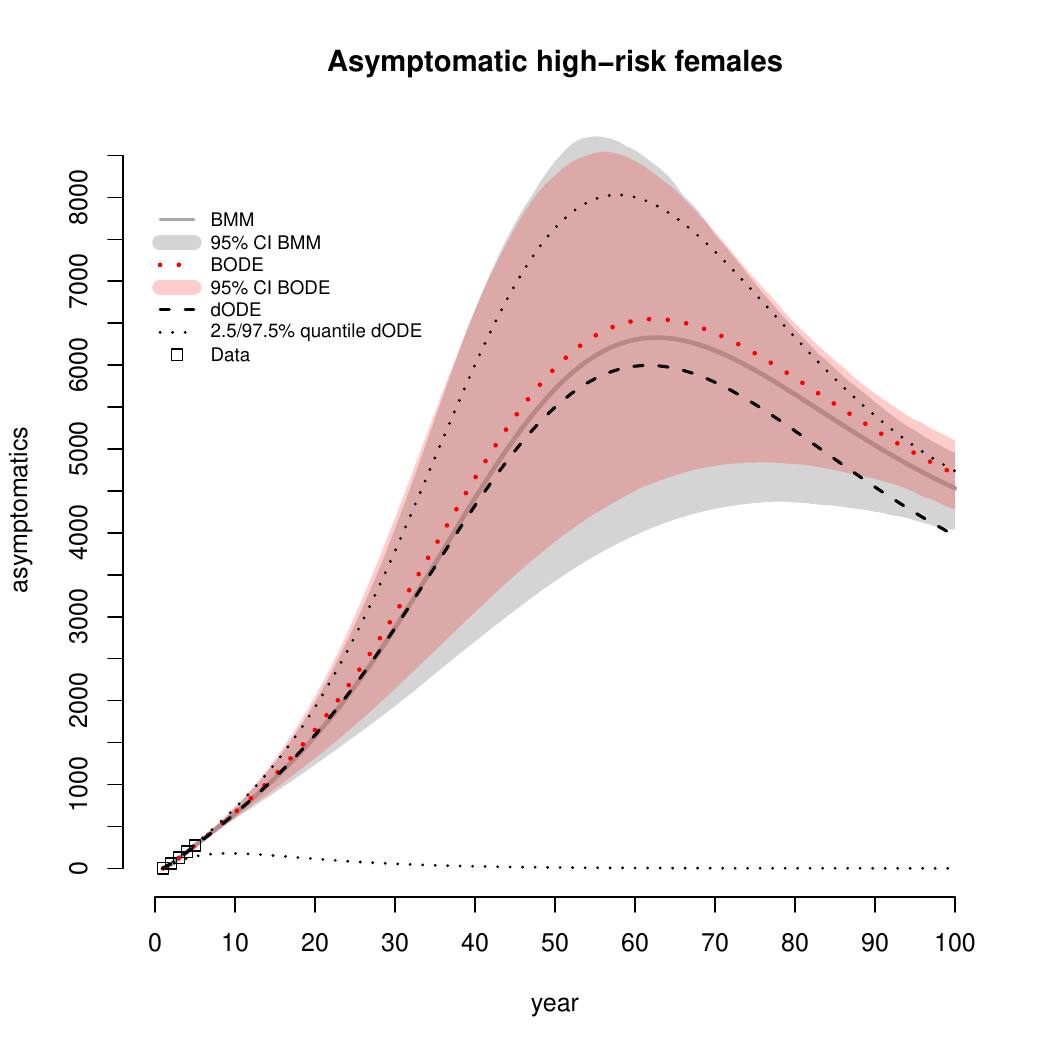}
\end{subfigure}
\begin{subfigure}{.5\textwidth}
\centering
\includegraphics[width=0.9\linewidth]{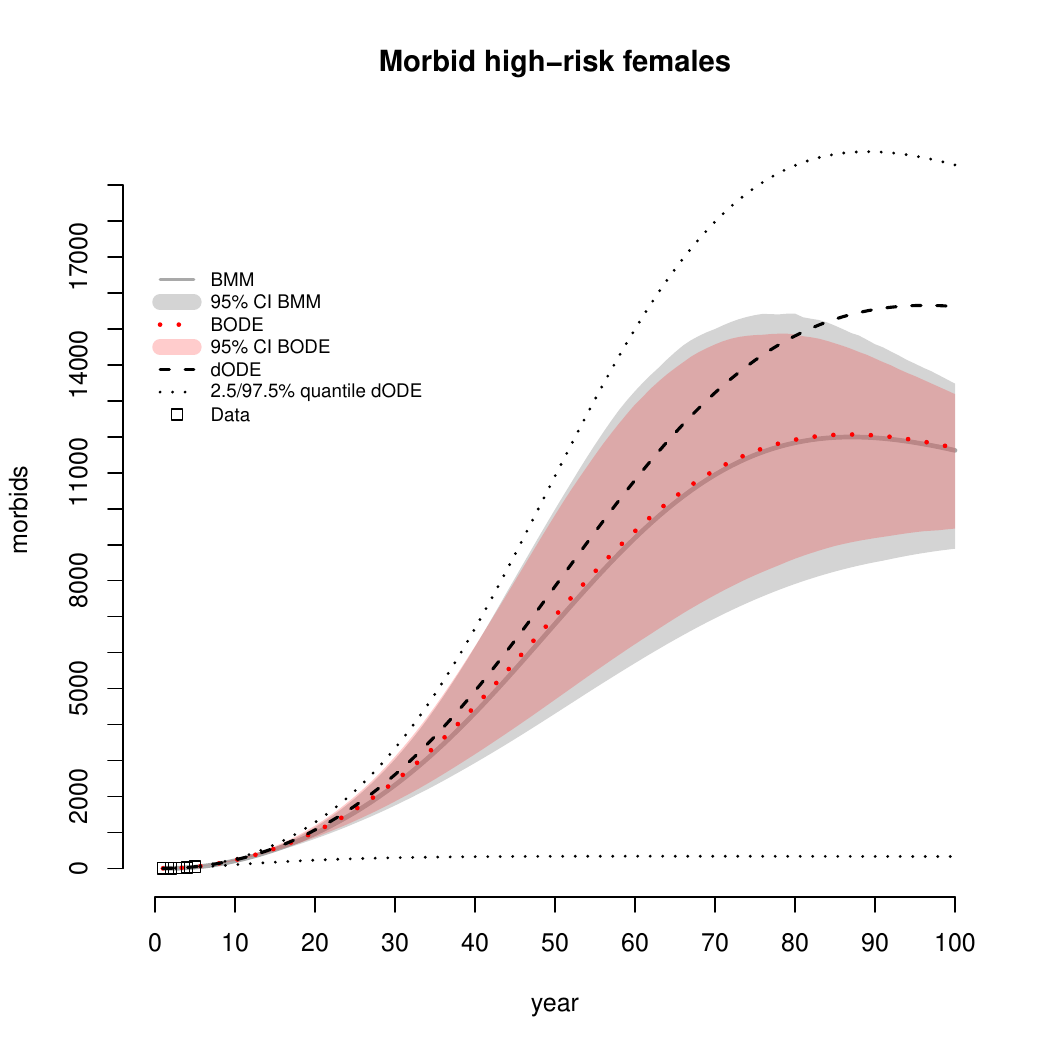}
\end{subfigure}
\caption{Calibration results on the number of high-risk females in the states following a systematic probabilistic calibration approach. The results of the Bayesian models are similar, with a slightly higher number of high-risk females in the states \textit{Infected} and \textit{Asymptomatic} estimated by the Bayesian ODE-based model. In contrast, the deterministic ODE-based model results in a lower estimate on the number of high-risk females in the states \textit{Infected} and \textit{Asymptomatic}; however, the outcome on the state \textit{Morbid} is~reversed.\label{fig:calstates}}
\end{figure}

Figure \ref{fig:calstates} shows the outcome on the natural history of the fictional chronic STI following calibration. Only the results on high-risk females are displayed; those on high-risk males are comparable. The BMM produces results which are comparable to the ``gold standard'' of the BODE. The model outcome on the states \textit{Susceptible} and \textit{Morbid} of the BMM and BODE is basically identical, whereas the outcome of the BODE shows slightly higher estimates on the number of infected and asymptomatic high-risk females. The outcome on the number of susceptible and morbid high-risk females is higher in the dODE; in contrast, the outcome on those in the states \textit{Infected} and \textit{Asymptomatic} is lower when compared to the two Bayesian models. The ranges of the 95\% credible intervals of the BODE and BMM are similar, showing wider ranges in the BMM. The 97.5\% quantiles of the scenario analysis of the dODE are lower than the upper bounds of the 95\% CIs in the states \textit{Infected} and \textit{Asymptomatic} and higher in the states \textit{Susceptible} and \textit{Morbid}, whereas the 2.5\% quantiles are considerably lower than the lower bounds of the CIs (apart from the state \textit{Susceptible}).

\subsection{Cost-effectiveness analysis\label{cea}}
We denote the unit costs and utilities as $c_{sti}$ and $u_s$, with indices $s$, $t$ and $i$ representing states \mbox{$s\in\{1,...,S\}$}, observation time points $t$ and interventions $i=1$ (status quo) and $i=2$ (vaccination). We assume decreasing utility values for more severe states. Costs are induced by screening, vaccination, a visit at the general practitioner and diagnostic tests. Following a positive STI diagnosis, further diagnostic tests and treatment are necessary. For all these quantities, the distributional assumptions are presented in Table \ref{tab:param}. 

The overall costs per intervention are calculated as 
\[C_i=\displaystyle\sum_{t=1}^T\sum_{s=1}^S\frac{c_{si}n_{sti}}{(1+\delta)^{t-1}},\] 
where $n_{sti}$ are the number of people in state $s$ at time $t$ when intervention $i$ is applied and $\delta$ is the yearly discount rate. In both the continuous- and discrete-time approaches, the model output on the natural history of disease infection and progression is evaluated at pre-specified time points $t \in \{1,...,T\}$, where $T$ represents the end of follow-up. We discount both costs and benefits at a fixed yearly rate $\delta=0.03$, following ISPOR recommendations \cite{ISPOR}. Similarly, the overall utilities are computed as 
\[U_i=\displaystyle\sum_{t=1}^T\sum_{s=1}^S\frac{u_{s}n_{sti}}{(1+\delta)^{t-1}}.\]
 
Overall costs and utilities define the monetary net benefit $\mbox{NB}_i(\bm{\theta})=kU_{i}-C_{i}$. The economic evaluation is performed by calculating suitable summaries such as the increment in mean cost $\Delta_c=C_{2}-C_{1}$ and the increment in mean effectiveness $\Delta_e=U_{2}-U_{1}$ between vaccination and the status-quo, or the incremental cost-effective\-ness ratio 
\[\mbox{ICER}=\frac{\mbox{E}[\Delta_c]}{\mbox{E}[\Delta_e]}.\] 
In the BMM and BODE, these quantities are estimated directly as function of the parameters, while in the dODE, we conduct a scenario analysis including the 2.5\% and 97.5\% quantiles of the ICER to evaluate the range of ``plausible'' results. A cut-off point of a \textit{willingness-to-pay} $k$ of approximately \pounds 20,000 -- \pounds 30,000 per QALY gained, adopted by NICE \cite{Rawlins}, is used as the benchmark of value for money. 

As for PSA, it is usually based on: (\textit{i}) the analysis of the cost-effectiveness plane, depicting the joint probability distribution of $(\Delta_e,\Delta_c)$; (\textit{ii}) the cost-effectiveness acceptability curve \mbox{$\mbox{CEAC}=\Pr(k\Delta_e-\Delta_c>0)$}, which shows the probability that the reference intervention is cost-effective as a function of the willingness to pay $k$; and (\textit{iii}) the expected value of ``perfect'' information
\[ \mbox{EVPI} =  \mbox{E}_{\bm\theta}\left[\max_i\mbox{NB}_i\left(\bm\theta\right)\right] - \max_i\mbox{E}_{\bm\theta} \left[\mbox{NB}_i\left(\bm\theta\right)\right], \]
which quantifies the maximum amount of money that the decision-maker should be willing to invest (\textit{e.g.} in a new study) in order to resolve parameter uncertainty and thus make a ``better'' decision. The Bayesian models can perform these analyses in a straightforward way, since these quantities are all functions of the model parameters and thus a full posterior distribution can be directly obtained.

The ICER of the dODE results in \textsterling \,7,203.416, ranging between \textsterling \,2,592.44 and \textsterling \,469,906 in the scenario analysis at the 2.5\% and 97.5\% quantiles. The ICERs of the Bayesian models are comparable to the dODE, resulting in \textsterling \,6,054.82 and \textsterling \,6,287.62 in the BODE and \mbox{BMM,~respectively.} 

\begin{figure}[H]
\centering
\includegraphics[width=0.95\textwidth]{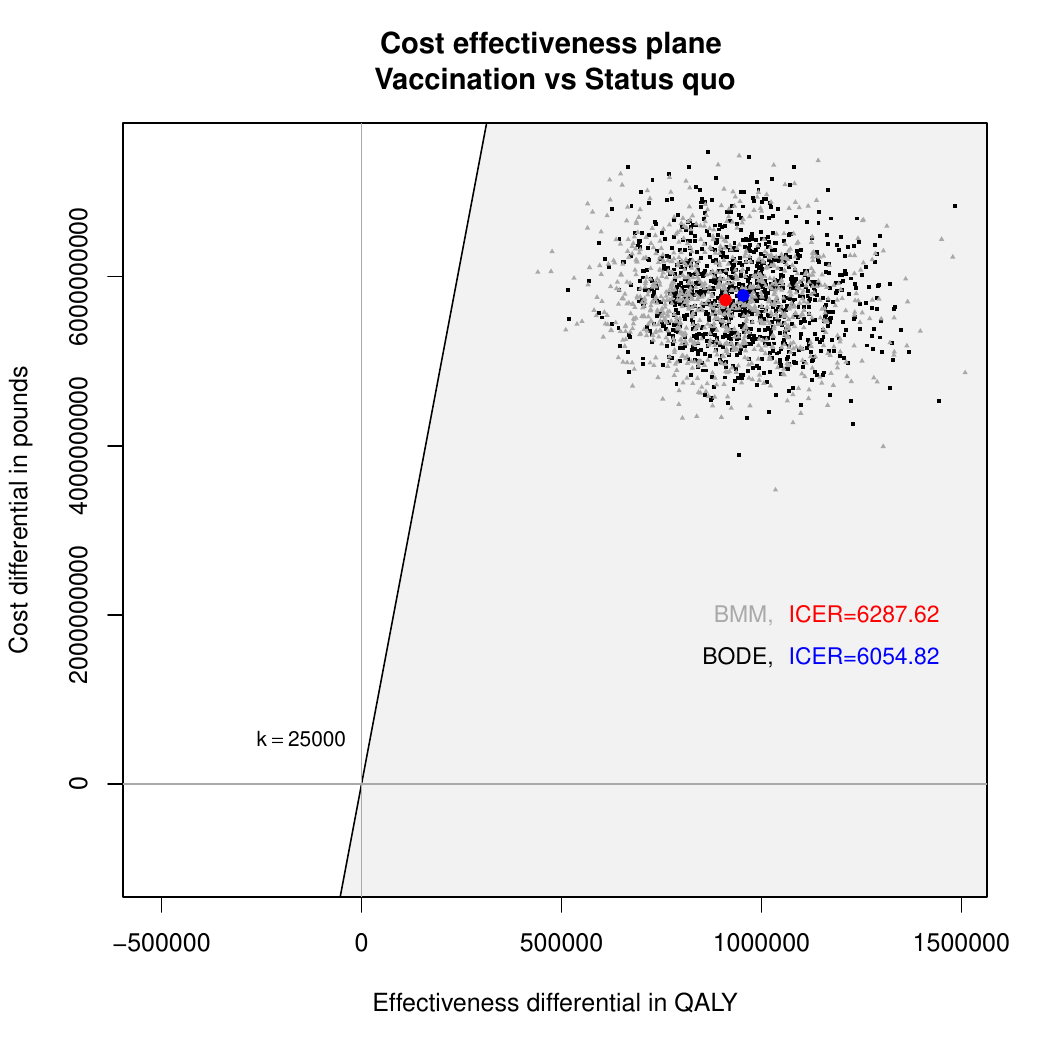}
\caption{Cost-effectiveness planes of the Bayesian ODE system and Bayesian Markov model. The cost-effectiveness plane indicates that vaccination is both more expensive and more effective than the status quo. All points lie within the sustainability area~of cost-effectiveness. The ICERs of \textsterling \,6,054.82 (blue dot, BODE) and \textsterling \,6,287.62 (red dot, BMM) indicate cost-effectiveness of STI vaccination in comparison to STI screening at a threshold of \textsterling \,25,000. \label{fig:ceaprobcal}}
\end{figure}

\mbox{Figure \ref{fig:ceaprobcal}} displays the cost-effectiveness plane. Each point of the MCMC simulation lies within the grey sustainability area, indicating that STI vaccination is cost-effective at a threshold of \textsterling \,25,000 when compared to STI screening. STI vaccination is deemed to be both more expensive and more effective than STI screening since all points are located in the upper right quadrant of the graph. The corresponding ICERs of the BMM and BODE are displayed as red and blue dots,~respectively.

\begin{figure}[H]
\centering
\begin{subfigure}{0.48\textwidth}
\centering
\includegraphics[width=\textwidth]{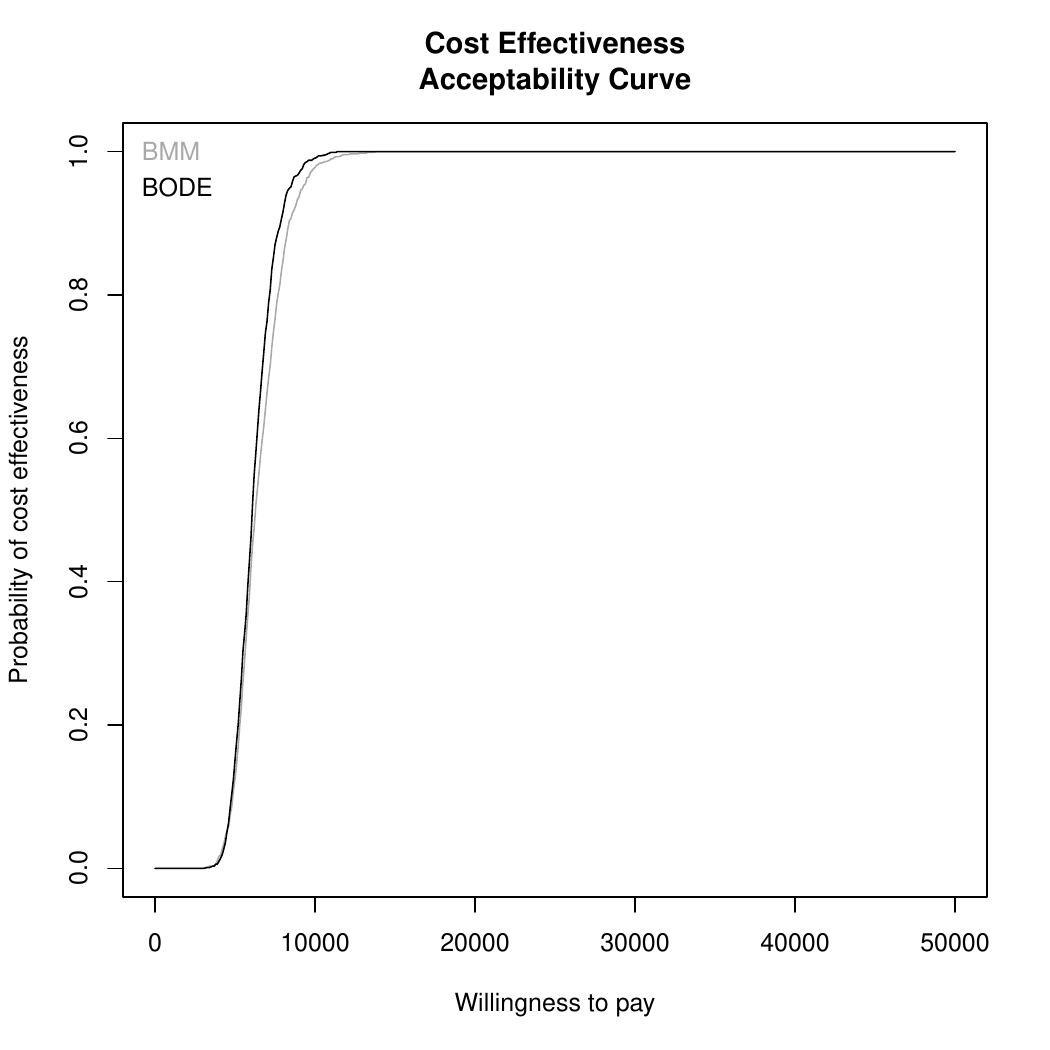}
\end{subfigure}
\begin{subfigure}{0.48\textwidth}
\centering
\includegraphics[width=\textwidth]{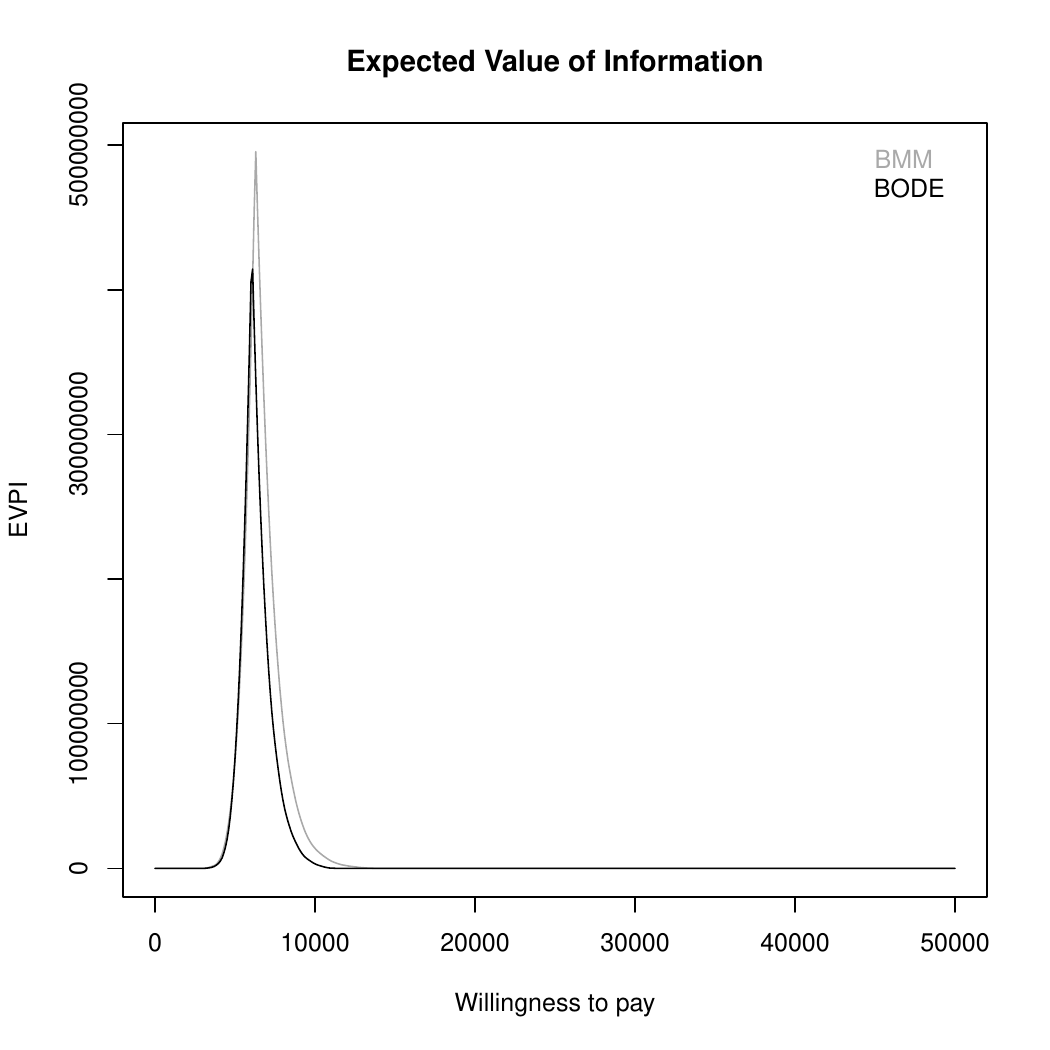}
\end{subfigure}
\caption{Cost-effectiveness acceptability curves and expected value of information of the
Bayesian ODE system and Bayesian Markov model. The results of the BMM are displayed in grey, whereas those of the BODE are shown in black. The amount of parameter uncertainty is higher in the BMM. The CEACs in the left panel reach values of 80\% at a willingness-to-pay corresponding to the ICERs. The EVPIs for the whole population at around \textsterling \,500,000,000 and \textsterling \,400,000,000 in the BMM and BODE, respectively, are shown in the right~panel.\label{fig:psacali}}
\end{figure}

Figure \ref{fig:psacali} shows the results of the CEACs and the population EVPIs of the two Bayesian models. The amount of uncertainty in the BMM is slightly larger than in the BODE; however, 80\% cost-effectiveness is clearly reached at the break-even points of the ICERs of both models. The population EVPI of the BMM at around \textsterling \,500,000,000 is higher than in the BODE, where it reaches a value of around \textsterling \,400,000,000. The higher EVPI value of the BMM is a consequence of the slightly larger amount of uncertainty. Thus, the value of additional research is higher in the~BMM.

\section{Discussion}
In this paper we have presented a comparison of modelling methods for the economic evaluation of interventions in infectious disease. We acknowledge that ODE-based models have several advantages and consider the Bayesian ODE structure as ideal to combine transmission modelling with economic evaluation. However, including a large number of states and probability distributions on a high number of model parameters induces computational issues, which effectively acts as a barrier to the application of complex economic modelling in this area. This possibly explains why the extensive application of PSA is limited in comparison to many other disease areas, in health~economics.

From the technical point of view, constructing a BMM is relatively simple and does not require the use of specialised software --- in fact our analysis has been performed using \texttt{R} and the Gibbs sampler \texttt{JAGS}, which are often used by statisticians and health economic modellers and thus by reviewers and advisers for health technology assessment agencies. This may again facilitate the communication of complex modelling assumptions and thus the economic assessment of complex interventions such as those based on vaccination programmes.

We of course acknowledge that the run time of the BODE could be reduced, for example using optimised code implemented in \texttt{C}. This is also true of the BMM, which could be coded using \texttt{C} directly. We note, however that a Bayesian implementation of an ODE system is more computationally intensive than the corresponding deterministic version, due to the MCMC component needed to obtain the estimate for the posterior distributions of the relevant parameters. In addition, our proposal aims at giving a general framework: it is possible that in specific cases modellers will be able to write highly optimised code that reduces the computational time. In general terms, however, our experience suggests that many modellers working in health economic evaluation would not be necessarily familiar with specialised code (e.g.~\texttt{C}).
 
Moreover, it is important that the code used to provide the evaluation of the transmission model is made available to reviewers (e.g. NICE appraisal committee), among whom a language such as \texttt{WinBUGS} is familiar, but other, more advanced programming languages are not. Finally, the possibility of estimating the transmission model and embedding in the economic model has the potential to improve the overall process (because the modellers developing the latter would be equipped with the technical expertise to develop the former too).

In addition to language, a more crucial point with respect to computational efficacy is given by the choice of the MCMC sampler. An alternative to a Gibbs sampler would for example be Hamiltonian Monte Carlo (HMC), which is a more efficient algorithm. Random walks are avoided using HMC since the gradient information is used. As a consequence, the algorithm is able to sample more efficiently from regions of high probability \cite{Brooks}.

\section{Conclusion}

Our proposal of a dynamic Bayesian Markov model can be seen as an effective compromise between the ideal ODE-based models including suitable distributions on all model parameters and simpler structures of MMs that fail to account for time-dependent changes in prevalence and the effects of herd immunity. While providing a sparser temporal resolution in the way in which transmission is modelled, our methodology has the advantage of accounting for parameter uncertainty. This in turn means that standard economic analysis, including PSA, can be performed in a straightforward way. In addition, Markov models are a well established tool in health economics, which may facilitate the translation of the modellers' work to the regulators and assessors.

In the fictional example presented in this paper, our BMM performs just as well as the BODE, with seizable computational savings. Model predictions should always be calibrated, given time series or prevalence data are available. Systematic calibration approaches usually induce a considerably high computational effort. We could show that including a direct calibration approach in the Bayesian models, the BMM runs 15 times faster than the BODE.


\glsaddall
\printglossary[title=List of abbreviations]

\section*{Declarations}

\subsection*{Competing interests}
The authors declare that they have no competing interests.
\subsection*{Funding}
Dr Gianluca Baio is partially supported as the recipient of a research grant sponsored by Mapi Group at University College London. The funding body was involved in the design of the study and in simulation, analysis and interpretation of data. The study, however, is based entirely on simulated data on a fictional chronic sexually transmitted infection.
\subsection*{Author's contributions}
Dr Katrin Haeussler conducted the literature review on methodology in dynamic infectious disease transmission modelling. Dr Katrin Haeussler and Dr Gianluca Baio contributed the approach of a dynamic Bayesian Markov model. Dr Katrin Haeussler wrote the corresponding programmes in \texttt{JAGS} and \texttt{WinBUGS}. Dr Gianluca Baio and Dr Ardo van den Hout provided helpful advice in selecting the calibration approaches as well as improving the methodology and the writing. All authors read and approved the final manuscript. 

\subsection*{Acknowledgements}
 Dr Sim\'{o}n Lunag\'{o}mez, lecturer at Lancaster University, provided thought-provoking and helpful advice in writing and revising the manuscript. His area of expertise is Bayesian modelling for network data with possible application to infectious disease. In addition, he is experienced in competing risk analysis.

\bibliographystyle{bmc-mathphys} 
\bibliography{Haeusslerbibliography}      

\begin{appendix}
\section*{Appendix}
\section{Sex- and behaviour-specific force of infection\label{appforce}}

In the case study described in Section~\ref{healthtrans}, the force of infection presented in (\ref{forceinf2}) is adjusted to additionally account for sex and behaviour. The indices $v,v'=$ ($M$ale, $F$emale) indicate the respective sex and its opposite. For example, a male is represented through the index $M$; the index of his female mixing partner is $F$. The sexual behaviour group is represented through the index $b=$ ($L$ow, $H$igh). In the ODE models, the transition rate $\rho_{v,b,1,2}(t)$ from the state \textit{Susceptible} to the state \textit{Infected} depends on the covariates sex and behaviour. In the BMM, the sex- and behavioural specific transition probability $\pi_{v,b,1,2,t}$ is estimated by transforming $\rho_{v,b,1,2}(t)$ according to (\ref{coop}).  

For simplicity, we exclusively account for random mixing; members of the population of sex $v$ randomly select sexual partners of the opposite sex $v'$. Because of the impact of the covariates, the estimation of the overall prevalence in the sexual partners of sex $v'$ is a weighted average of the prevalence in both behaviour groups of sex $v'$. We show the corresponding equations for the continuous-time approach as functions of $t$; for a discrete-time approach, these are similar. 

The time-specific probability of selecting a partner from the high-risk group, which we indicate as $g_{v'H}(t)$, depends on the partner acquisition rates $\omega_{v'H}$ and $\omega_{v'L}$ as well as on the population sizes $N_{v'H}(t)$ and $N_{v'L}(t)$. The probability of selecting a partner from the low-risk group is represented by $g_{v'L}(t)$. The corresponding equations adapted from \cite{Vynnycky} only account for two sexual behaviour groups and are thus extended for sex, to give

\begin{eqnarray*}\label{gh}
g_{v'H}(t)& = &\frac{\omega_{v'H}N_{v'H}(t)}{\omega_{v'H}N_{v'H}(t)+\omega_{v'L}N_{v'L}(t)} \\
g_{v'L}(t)& = & 1-g_{v'H}(t).
\end{eqnarray*}
We estimate the sex-, behavioural- and time-specific force of infection 
\begin{equation*}\label{force}
\rho_{v,b,1,2}(t)= \beta \omega_{vb}\overline{\psi}_{v'}(t),
\end{equation*}
where \[\overline{\psi}_{v'}(t)=\left(g_{v'H}(t)\frac{I_{v'H}(t)}{N_{v'H}(t)}+g_{v'L}(t)\frac{I_{v'L}(t)}{N_{v'L}(t)}\right)\] is the weighted average of the STI population prevalence, which is estimated as a function of the probabilities $g_{v'b}(t)$ of selecting a partner of the opposite sex from one of the two sexual behaviour groups and the time-, sex- and behavioural-specific population prevalence $\frac{I_{v'b}(t)}{N_{v'b}(t)}$. The number of infectious people $I_{v'b}(t)$ is estimated as those in the state \textit{Infected} of the respective sex and behaviour group. In line with (\ref{forceinf2}), the force of infection $\rho_{v,b,1,2}(t)$ is a function of the STI transmission probability per partnership $\beta$, the partner acquisition rates $\omega_{vb}$, and the population prevalence $\overline{\psi}_{v'}(t)$.

\section{Model calibration\label{cal}}

The dODE is calibrated through a frequentist probabilistic calibration approach \cite{Vanni,vandeVelde3}. Suitable distributions are assigned to the model parameters which are relevant in terms of natural history of disease. As a next step, sets of 50,000 parameter combinations are sampled from the distributions through Monte Carlo sampling. We calculate the sum of squared errors between the outputs of the model runs (for each parameter set combination) and simulated data for the first five years of follow-up in four states~as

\begin{equation*}
\mathcal{Q}(\bm{\theta})=\sum_{t=1}^5\sum_{s=1}^4\left[y_s(t)-f_s\left(t\mid\bm{\theta}\right)\right]^2.
\end{equation*}

The dODE is based on a continuous-time approach; however, the corresponding output is evaluated at yearly time intervals $t\in\{1,...,5\}$. The simulated data on the number of high-risk people in state $s$ at time $t$ are indicated by $y_s(t)$, and $f_s(t\mid\bm{\theta})$ is the model output on high-risk people given the input parameter set $\bm{\theta}$. As a final step, the set $\bm{\theta}$ corresponding to the output which results in the least sum of squares is selected; it is displayed in Table \ref{tab:dodePar}. 

\begin{table}[H]
\fontsize{7.5}{6.5}\selectfont
\caption[Point estimates of the parameters of the deterministic ODE-based model obtained through a frequentist probabilistic calibration approach]{Point estimates of the parameters of the deterministic ODE-based model obtained through a frequentist probabilistic calibration approach. The parameter set $\bm{\theta}$ with the best fit to simulated data minimises the sum of squared~errors. \label{tab:dodePar}}
\centering
\begin{tabular}{llc}
\hline
\textbf{Parameter}&\textbf{Description}&\textbf{Point estimate}\\
\hline
$\omega_{M\!H}$&Partner acquisition rate (high-risk males)& 8.3515\\
$\omega_{M\!L}$&Partner acquisition rate (low-risk males)& 2.4526 \\
$\omega_{F\!H}$&Partner acquisition rate (high-risk females)& 8.3836\\
$\omega_{F\!L}$&Partner acquisition rate (low-risk females)& 1.6085\\
$\chi$&Proliferation parameter & 0.0100\\
$\beta$&STI transmission probability per partnership& 0.1639\\
$\pi_{\mbox{\tiny{2,3}}}$&Transition parameter from state $2$ to state $3$& 0.7957\\
$\pi_{\mbox{\tiny{3,4}}}$&Transition parameter from state $3$ to state $4$& 0.0891\\
$\pi_{\mbox{\tiny{4,5}}}$&Transition parameter from state $4$ to state $5$& 0.0232\\
$\pi_{\mbox{\tiny{1,5}}}$&Transition parameter from state $1$ to state $5$& 0.0005\\
\hline
\end{tabular}
\end{table}

In contrast to the dODE, the BMM and BODE are calibrated through a Bayesian calibration approach \cite{deAngelis,Welton2}. As described in Section~\ref{accparun}, the advantage is that model parameters can be inferred by fitting the models to data directly (in one step). 

In a Bayesian framework, data can be considered in several ways. We include simulated data on a selection of parameters to update the priors into the corresponding posteriors. However, this only ensures that the corresponding parameters are informed by available evidence. Despite posterior sampling, it could still be possible that the predicted outcome implied by the Bayesian models was not comparable to high-quality data obtained from large data registries. In the BODE, the output on the number of people in the states over follow-up corresponds to the solutions of the ODEs. In the BMM, the output corresponds to the solutions of the state allocation algorithm, which is given by (\ref{allocation}). In addition to posterior sampling, the outputs of both models are constantly updated through the simulated time series data. In fact, this is an additional process of Bayesian inference, updating the model outcome through additional data. 

The updating takes place by assigning Poisson distributions to the data; the event rate $\lambda$ is then represented by the solutions of the ODEs and the state allocation algorithm, respectively. The resulting model outcome is already calibrated, and no further steps such as the calculation of goodness-of-fit statistics are necessary. The evaluation of the natural history of disease and the cost-effectiveness analysis including PSA can be conducted~straightforwardly.

\section{Code of the case study\label{Rcode}}

\subsection{The Bayesian ODE-based model}

\begin{lstlisting}
model{

#define ODEs in WBDiff syntax
solution[1:n.grid,1:dim,1:2]<-ode(init[1:dim,1:2],grid[1:n.grid],D(y[1:dim,1:2],t),origin,tol)

#ODEs for interventions i=1 (screening only) and i=2 (vaccination)
for(i in 1:2){

	#ODEs for low-risk females
	#Susceptible
	D(y[1,i],t) <-chi*nfemL[i]-lambdafemL[i]*y[1,i]-rhosd*y[1,i]

	#Infected
	D(y[2,i],t) <-lambdafemL[i]*y[1,i]-rhoia*y[2,i]-rhosd*y[2,i]

	#Asymptomatic
	D(y[3,i],t) <-rhoia*y[2,i]-rhoam*y[3,i]-rhosd*y[3,i]

	#Morbid
	D(y[4,i],t) <-rhoam*y[3,i]-rhomd*y[4,i]-rhosd*y[4,i]

	#Dead
	D(y[5,i],t) <-rhosd*nfemL[i]+rhomd*y[4,i]

	#ODEs for high-risk females
	#Susceptible
	D(y[6,i],t) <-chi*nfemH[i]-lambdafemH[i]*y[6,i]-rhosd*y[6,i]

	#Infected
	D(y[7,i],t) <-lambdafemH[i]*y[6,i]-rhoia*y[7,i]-rhosd*y[7,i]

	#Asymptomatic
	D(y[8,i],t) <-rhoia*y[7,i]-rhoam*y[8,i]-rhosd*y[8,i]

	#Morbid
	D(y[9,i],t) <-rhoam*y[8,i]-rhomd*y[9,i]-rhosd*y[9,i]

	#Dead
	D(y[10,i],t) <-rhosd*nfemH[i]+rhomd*y[9,i]

	#ODEs for low-risk males
	#Susceptible
	D(y[11,i],t) <-chi*nmalL[i]-lambdamalL[i]*y[11,i]-rhosd*y[11,i]

	#Infected
	D(y[12,i],t) <-lambdamalL[i]*y[11,i]-rhoia*y[12,i]-rhosd*y[12,i]

	#Asymptomatic
	D(y[13,i],t) <-rhoia*y[12,i]-rhoam*y[13,i]-rhosd*y[13,i]

	#Morbid
	D(y[14,i],t) <-rhoam*y[13,i]-rhomd*y[14,i]-rhosd*y[14,i]

	#Dead
	D(y[15,i],t) <-rhosd*nmalL[i]+rhomd*y[14,i]

	#ODEs for high-risk males
	#Susceptible
	D(y[16,i],t) <-chi*nmalH[i]-lambdamalH[i]*y[16,i]-rhosd*y[16,i]

	#Infected
	D(y[17,i],t) <-lambdamalH[i]*y[16,i]-rhoia*y[17,i]-rhosd*y[17,i]

	#Asymptomatic
	D(y[18,i],t) <-rhoia*y[17,i]-rhoam*y[18,i]-rhosd*y[18,i]

	#Morbid
	D(y[19,i],t) <-rhoam*y[18,i]-rhomd*y[19,i]-rhosd*y[19,i]

	#Dead
	D(y[20,i],t) <-rhosd*nmalH[i]+rhomd*y[19,i]

	#number of low-risk females alive
	nfemL[i] <- y[1,i]+y[2,i]+y[3,i]+y[4,i]

	#number of high-risk females alive
	nfemH[i] <- y[6,i]+y[7,i]+y[8,i]+y[9,i]

	#number of low-risk males alive
	nmalL[i]<-y[11,i]+y[12,i]+y[13,i]+y[14,i]

	#number of high-risk males alive
	nmalH[i]<-y[16,i]+y[17,i]+y[18,i]+y[19,i]

	#probability of selecting a high-risk male partner
	gmalH[i] <- (rmalH*nmalH[i])/(rmalH*nmalH[i]+rmalL*nmalL[i])

	#probability of selecting a low-risk male partner
	gmalL[i] <- 1-gmalH[i]

	#probability of selecting a high-risk female partner
	gfemH[i] <- (rfemH*nfemH[i])/(rfemH*nfemH[i]+rfemL*nfemL[i])

	#probability of selecting a low-risk female partner
	gfemL[i] <- 1-gfemH[i]

	#prevalence in males
	pmal[i] <- gmalH[i]*(y[17,i]/nmalH[i])+gmalL[i]*(y[12,i]/nmalL[i])

	#prevalence in females
	pfem[i] <- gfemH[i]*(y[7,i]/nfemH[i])+gfemL[i]*(y[2,i]/nfemL[i])

	#intervention screening: vaccine coverage (pi2=1), vaccine efficacy (eff2=0)
	#intervention vaccination: vaccine coverage (pi2=pi), vaccine efficacy (eff2=eff)
	pi2[i]<-pi*step(i-1.1)+(1-step(i-1.1))
	eff2[i]<-eff*step(i-1.1)

	#force of infection in the two risk groups and sexes
	lambdafemL[i] <-pi2[i]*(1-eff2[i])*beta*rfemL*pmal[i]+(1-pi2[i])*beta*rfemL*pmal[i]
	lambdafemH[i] <-pi2[i]*(1-eff2[i])*beta*rfemH*pmal[i]+(1-pi2[i])*beta*rfemH*pmal[i]
	lambdamalL[i] <-pi2[i]*(1-eff2[i])*beta*rmalL*pfem[i]+(1-pi2[i])*beta*rmalL*pfem[i]
	lambdamalH[i] <-pi2[i]*(1-eff2[i])*beta*rmalH*pfem[i]+(1-pi2[i])*beta*rmalH*pfem[i]
}

#Initial conditions intervention screening-only:
init[1,1] <- 399760; init[2,1] <- 240; init[3,1] <- 0; init[4,1] <- 0; init[5,1] <- 0; 
init[6,1] <- 99940; init[7,1] <- 60; init[8,1] <- 0; init[9,1] <- 0; init[10,1] <- 0; 
init[11,1] <- 399760; init[12,1] <- 240; init[13,1] <- 0; init[14,1] <- 0; 
init[15,1] <- 0; init[16,1] <-  99940; init[17,1] <- 60; init[18,1]<- 0; 
init[19,1] <- 0; init[20,1] <- 0

#initial conditions intervention vaccination are equivalent
init[1,2]<-init[1,1]; init[2,2]<-init[2,1]; init[3,2]<-init[3,1]; init[4,2]<-init[4,1]; 
init[5,2]<-init[5,1]; init[6,2]<-init[6,1]; init[7,2]<-init[7,1]; init[8,2]<-init[8,1]; 
init[9,2]<-init[9,1]; init[10,2]<-init[10,1]; init[11,2]<-init[11,1]; 
init[12,2]<-init[12,1]; init[13,2]<-init[13,1]; init[14,2]<-init[14,1]; 
init[15,2]<-init[15,1]; init[16,2]<-init[16,1]; init[17,2]<-init[17,1]; 
init[18,2]<-init[18,1]; init[19,2]<-init[19,1]; init[20,2]<-init[20,1]

#poisson-gamma model on partner acquisition rates
for(i in 1:popsize) {
	datfemH[i] ~ dpois(rfemH)
	datfemL[i] ~ dpois(rfemL)
	datmalH[i] ~ dpois(rmalH)
	datmalL[i] ~ dpois(rmalL)
}

#informative priors on partner acquisition rates
rfemH ~ dgamma(0.1,0.1)
rfemL ~ dgamma(0.1,0.1)
rmalH ~ dgamma(0.1,0.1)
rmalL ~ dgamma(0.1,0.1)

#informative prior on reproduction rate
chi~dgamma(1111.1,111111.1)

#informative prior on transition rate infected --> asymptomatic
rhoia~dgamma(25600,32000)

#informative prior on transition rate asymptomatic --> morbid
rhoam~dgamma(2025,22500)

#informative prior on transition rate morbid --> dead
rhomd~dgamma(1600,40000)

#informative prior on transition rate susceptible --> dead
rhosd~dgamma(156.25,312500)

#beta-binomial model on STI transmission probability per partnership. r.beta: number of events, n.beta: sample size
r.beta ~ dbin(beta,n.beta)
beta ~ dbeta(0.5,0.5)

#beta-binomial model on vaccine coverage
r.pi ~ dbin(pi,n.pi)
pi ~ dbeta(0.5,0.5)

#beta-binomial model on vaccine efficacy
r.eff ~ dbin(eff,n.eff)
eff ~ dbeta(0.5,0.5)

#beta-binomial model on probability of diagnosis
r.diag ~ dbin(diag,n.diag)
diag ~ dbeta(0.5,0.5)

#beta-binomial model on probability of screening
r.screen ~ dbin(screen,n.screen)
screen ~ dbeta(0.5,0.5)

#Bayesian model calibration through data on the number of high-risk females and males in the states susceptible, infected, asymptomatic and morbid. These are count data and therefore modelled through poisson distributions. This is done for the first time point of follow-up.

dat.sus.femH[1] ~ dpois(init[6,1])
dat.inf.femH[1] ~ dpois(init[7,1])
dat.asy.femH[1] ~ dpois(init[8,1])
dat.mor.femH[1] ~ dpois(init[9,1])
dat.sus.malH[1] ~ dpois(init[16,1])
dat.inf.malH[1] ~ dpois(init[17,1])
dat.asy.malH[1] ~ dpois(init[18,1])
dat.mor.malH[1] ~ dpois(init[19,1])

#Data for time points 2-5 are also available and used to calibrate high-risk people in four of the states. The constraint (maximum) is included due to a computational issue, is not relevant and can be omitted.

for (n in 2:5){
    conssusfemH[n] <- max(0.1,solution[n,6,1])
    dat.sus.femH[n] ~ dpois(conssusfemH[n])
    consinffemH[n] <- max(0.1,solution[n,7,1])
    dat.inf.femH[n] ~ dpois(consinffemH[n])
    consasyfemH[n] <- max(0.1,solution[n,8,1])
    dat.asy.femH[n] ~ dpois(consasyfemH[n])
    consmorfemH[n] <- max(0.1,solution[n,9,1])
    dat.mor.femH[n] ~ dpois(consmorfemH[n])
    conssusmalH[n] <- max(0.1,solution[n,16,1])
    dat.sus.malH[n] ~ dpois(conssusmalH[n])
    consinfmalH[n] <- max(0.1,solution[n,17,1])
    dat.inf.malH[n] ~ dpois(consinfmalH[n])
    consasymalH[n] <- max(0.1,solution[n,18,1])
    dat.asy.malH[n] ~ dpois(consasymalH[n])
    consmormalH[n] <- max(0.1,solution[n,19,1])
    dat.mor.malH[n] ~ dpois(consmormalH[n])
}

#informative priors on costs and utilities 
tau.screen<-1/pow(0.693,2)
c.screen ~ dlnorm(2.996,tau.screen)
tau.vac<-1/pow(0.07986607,2)
c.vac ~ dlnorm(5.518352,tau.vac)
tau.test.sti<-1/pow(0.03,2)
c.test.sti ~ dlnorm(2.996,tau.test.sti)
tau.test.blood<-1/pow(0.03,2)
c.test.blood ~ dlnorm(3.401,tau.test.blood)
tau.trt<-1/pow(0.8325546,2)
c.trt ~ dlnorm(4.258597,tau.trt)
tau.trt.dis<-1/pow(0.1980422,2)
c.trt.dis ~ dlnorm(6.194998,tau.trt.dis)
tau.gp<-1/pow(0.02,2)
c.gp ~ dlnorm(3.912,tau.gp)
u.inf ~ dbeta(1469.3,629.7)
u.asym ~ dbeta(1439.4,959.6)
u.morb ~ dbeta(629.7,1469.3)
}
\end{lstlisting}

\subsection{The Bayesian Markov model}

\subsection*{\texttt{JAGS} model}
\begin{lstlisting}
###JAGS model
model {
#y: array including the number of people per year in the states. The first dimension is the year of follow-up, whereas the second dimension is the state with 1=susceptible low-risk females, 2=infected low-risk females, 3=asymptomatic low-risk females, 
		4=morbid low-risk females, 5=dead low-risk females, 6=susceptible high-risk females, 
		7=infected high-risk females, 8=asymptomatic high-risk females, 9=morbid high-risk 
		females, 10=dead high-risk females, 11=susceptible low-risk males, 12=infected low-
		risk males, 13=asymptomatic low-risk males, 14=morbid low-risk males, 15=dead low-
		risk males, 16=susceptible high-risk males, 17=infected high-risk males, 18=asympto-
		matic high-risk males, 19=morbid high-risk males, 20=dead high-risk males. The third 
		dimension is the intervention (1=screening, 2=vaccination).

#initialization of states (5 states * 2 sexes * 2 risk groups = 20 entries) for year 1 of follow-up (first dimension of y)

y[1,1,1] <- 399760; y[1,2,1] <- 240; y[1,3,1] <- 0; y[1,4,1] <- 0; y[1,5,1] <- 0; 
y[1,6,1] <- 99940; y[1,7,1] <- 60; y[1,8,1] <- 0; y[1,9,1] <- 0; y[1,10,1] <- 0; 
y[1,11,1] <- 399760; y[1,12,1] <- 240; y[1,13,1] <- 0; y[1,14,1] <- 0; 
y[1,15,1] <- 0; y[1,16,1] <- 99940; y[1,17,1] <- 60; y[1,18,1] <- 0; y[1,19,1] <- 0; 
y[1,20,1] <-0; y[1,1,2] <- y[1,1,1]; y[1,2,2] <- y[1,2,1]; y[1,3,2] <- y[1,3,1]; 
y[1,4,2] <- y[1,4,1]; y[1,5,2] <- y[1,5,1]; y[1,6,2] <- y[1,6,1]; y[1,7,2] <- y[1,7,1]; 
y[1,8,2] <- y[1,8,1]; y[1,9,2] <- y[1,9,1]; y[1,10,2] <- y[1,10,1]; 
y[1,11,2] <- y[1,11,1]; y[1,12,2] <- y[1,12,1]; y[1,13,2] <- y[1,13,1]; 
y[1,14,2] <- y[1,14,1]; y[1,15,2] <- y[1,15,1]; y[1,16,2] <- y[1,16,1]; 
y[1,17,2] <- y[1,17,1]; y[1,18,2] <- y[1,18,1]; y[1,19,2] <- y[1,19,1]; 
y[1,20,2] <-y[1,20,1];

#calculation of sample size according to risk-group and sex for year 1 of follow-up; only those alive are considered excluding those in the state of death with indices 5, 10, 15, 20)

nfemL[1,1] <- y[1,1,1]+y[1,2,1]+y[1,3,1]+y[1,4,1]
nfemH[1,1] <- y[1,6,1]+y[1,7,1]+y[1,8,1]+y[1,9,1]
nmalL[1,1] <- y[1,11,1]+y[1,12,1]+y[1,13,1]+y[1,14,1]
nmalH[1,1] <- y[1,16,1]+y[1,17,1]+y[1,18,1]+y[1,19,1]
nfemL[1,2]<-nfemL[1,1]
nfemH[1,2]<-nfemH[1,1]
nmalL[1,2]<-nmalL[1,1]
nmalH[1,2]<-nmalH[1,1]

#the probabilities of selecting a male/female partner from the high- or low-risk group, respectively, in year 1 of follow-up

gmalH[1,1] <- (rmalH*nmalH[1,1])/(rmalH*nmalH[1,1]+rmalL*nmalL[1,1])
gmalL[1,1] <- 1-gmalH[1,1]
gfemH[1,1] <- (rfemH*nfemH[1,1])/(rfemH*nfemH[1,1]+rfemL*nfemL[1,1])
gfemL[1,1] <- 1-gfemH[1,1]

gmalH[1,2]<-gmalH[1,1]
gmalL[1,2]<-gmalL[1,1]
gfemH[1,2]<-gfemH[1,1]
gfemL[1,2]<-gfemL[1,1]

#weighted average of STI prevalence in males in year 1 of follow-up 
pmal[1,1] <- gmalH[1,1]*(y[1,17,1]/nmalH[1,1])+gmalL[1,1]*(y[1,12,1]/nmalL[1,1])

#weighted average of STI prevalence in females in year 1 of follow-up
pfem[1,1] <- gfemH[1,1]*(y[1,7,1]/nfemH[1,1])+gfemL[1,1]*(y[1,2,1]/nfemL[1,1])

pmal[1,2]<-pmal[1,1]
pfem[1,2]<-pfem[1,1]

#force of infection: function of STI transmission probability beta, partner acquisition rates in low-risk females rfemL (accordingly for the other sex and risk group), and sex-specific population prevalence pmal and pfem, in year 1 of follow-up)

lambdafemL[1,1] <-1-exp(-beta*rfemL*pmal[1,1])
lambdafemH[1,1] <-1-exp(-beta*rfemH*pmal[1,1])
lambdamalL[1,1] <-1-exp(-beta*rmalL*pfem[1,1])
lambdamalH[1,1] <-1-exp(-beta*rmalH*pfem[1,1])
lambdafemL[1,2] <-1-exp(-(pi*(1-eff)*beta*rfemL*pmal[1,2]+(1-pi)*beta*rfemL*pmal[1,2]))
lambdafemH[1,2] <-1-exp(-(pi*(1-eff)*beta*rfemH*pmal[1,2]+(1-pi)*beta*rfemH*pmal[1,2]))
lambdamalL[1,2] <-1-exp(-(pi*(1-eff)*beta*rmalL*pfem[1,2]+(1-pi)*beta*rmalL*pfem[1,2]))
lambdamalH[1,2] <-1-exp(-(pi*(1-eff)*beta*rmalH*pfem[1,2]+(1-pi)*beta*rmalH*pfem[1,2]))

#dat.sus.femH: simulated data on susceptible high-risk females, used for calibration in year 1. The parameter y[1,6,1] contains the number of susceptible high-risk females in year 1 and is calibrated by the data for year 1. The approach for high-risk females in the states infected, asymptomatic and morbid as well as for high-risk males in the same states is conducted accordingly.

dat.sus.femH[1] ~ dpois(y[1,6,1])
dat.inf.femH[1] ~ dpois(y[1,7,1])
dat.asy.femH[1] ~ dpois(y[1,8,1])
dat.mor.femH[1] ~ dpois(y[1,9,1])  
dat.sus.malH[1] ~ dpois(y[1,16,1])
dat.inf.malH[1] ~ dpois(y[1,17,1])
dat.asy.malH[1] ~ dpois(y[1,18,1])
dat.mor.malH[1] ~ dpois(y[1,19,1])

#calibration in years 2-5
for(a in 2:5){
	dat.sus.femH[a] ~ dpois(max(0.1,y[a,6,1]))
	dat.inf.femH[a] ~ dpois(max(0.1,y[a,7,1]))
	dat.asy.femH[a] ~ dpois(max(0.1,y[a,8,1]))
	dat.mor.femH[a] ~ dpois(max(0.1,y[a,9,1]))   
	dat.sus.malH[a] ~ dpois(max(0.1,y[a,16,1]))
	dat.inf.malH[a] ~ dpois(max(0.1,y[a,17,1]))
	dat.asy.malH[a] ~ dpois(max(0.1,y[a,18,1]))
	dat.mor.malH[a] ~ dpois(max(0.1,y[a,19,1]))
}

for (i in 1:2){
	#intervention screening-only: pi2=1, else: pi2=pi
	pi2[i]<-ifelse(i==1,1,pi)
	#intervention screening-only: eff2=0 (no vaccine efficacy), else: eff2=eff
	eff2[i]<-ifelse(i==1,0,eff)
	###state allocation algorithm
	for (a in 2:A){
		#chi is the reproduction rate
		#nfemL and lambdafemL are explained above
		#pisd is the probability of death from any cause
		#piia is the transition probability infection --> asymptomatic
		#piam is the transition probability asymptomatic --> morbid
		#pimd is the transition probability morbid --> dead
		
		#state allocation of low-risk females in year a; y[a,1,i]=susceptible, 
				y[a,2,i]=infected, y[a,3,i]=asymptomatic, y[a,4,i]=morbid, y[a,5,i]=dead
		
		y[a,1,i] <- y[a-1,1,i]+chi*nfemL[a-1,i]-lambdafemL[a-1,i]*y[a-1,1,i]-pisd*y[a-1,1,i]
		y[a,2,i] <- y[a-1,2,i]+lambdafemL[a-1,i]*y[a-1,1,i]-piia*y[a-1,2,i]-pisd*y[a-1,2,i]
		y[a,3,i] <- y[a-1,3,i]+piia*y[a-1,2,i]-piam*y[a-1,3,i]-pisd*y[a-1,3,i]
		y[a,4,i] <- y[a-1,4,i]+piam*y[a-1,3,i]-pimd*y[a-1,4,i]-pisd*y[a-1,4,i]
		y[a,5,i] <- y[a-1,5,i]+pisd*nfemL[a-1,i]+pimd*y[a-1,4,i]
		
		#state allocation of high-risk females in year a; y[a,6,i]=susceptible, 
				y[a,7,i]=infected, y[a,8,i]=asymptomatic, y[a,9,i]=morbid, y[a,10,i]=dead
		
		y[a,6,i] <- y[a-1,6,i]+chi*nfemH[a-1,i]-lambdafemH[a-1,i]*y[a-1,6,i]-pisd*y[a-1,6,i]
		y[a,7,i] <- y[a-1,7,i]+lambdafemH[a-1,i]*y[a-1,6,i]-piia*y[a-1,7,i]-pisd*y[a-1,7,i]
		y[a,8,i] <- y[a-1,8,i]+piia*y[a-1,7,i]-piam*y[a-1,8,i]-pisd*y[a-1,8,i]
		y[a,9,i] <- y[a-1,9,i]+piam*y[a-1,8,i]-pimd*y[a-1,9,i]-pisd*y[a-1,9,i]
		y[a,10,i] <- y[a-1,10,i]+pisd*nfemH[a-1,i]+pimd*y[a-1,9,i]

		#state allocation of low-risk males in year a; y[a,11,i]=susceptible, 
				y[a,12,i]=infected, y[a,13,i]=asymptomatic, y[a,14,i]=morbid, y[a,15,i]=dead

		y[a,11,i] <- y[a-1,11,i]+chi*nmalL[a-1,i]-lambdamalL[a-1,i]*y[a-1,11,i]
				-pisd*y[a-1,11,i]
		y[a,12,i] <- y[a-1,12,i]+lambdamalL[a-1,i]*y[a-1,11,i]-piia*y[a-1,12,i]
				-pisd*y[a-1,12,i]
		y[a,13,i] <- y[a-1,13,i]+piia*y[a-1,12,i]-piam*y[a-1,13,i]-pisd*y[a-1,13,i]
		y[a,14,i] <- y[a-1,14,i]+piam*y[a-1,13,i]-pimd*y[a-1,14,i]-pisd*y[a-1,14,i]
		y[a,15,i] <- y[a-1,15,i]+pisd*nmalL[a-1,i]+pimd*y[a-1,14,i]

		#state allocation of high-risk males in year a; y[a,16,i]=susceptible, 
				y[a,17,i]=infected, y[a,18,i]=asymptomatic, y[a,19,i]=morbid, y[a,20,i]=dead

		y[a,16,i] <- y[a-1,16,i]+chi*nmalH[a-1,i]-lambdamalH[a-1,i]*y[a-1,16,i]
				-pisd*y[a-1,16,i]
		y[a,17,i] <- y[a-1,17,i]+lambdamalH[a-1,i]*y[a-1,16,i]-piia*y[a-1,17,i]
				-pisd*y[a-1,17,i]
		y[a,18,i] <- y[a-1,18,i]+piia*y[a-1,17,i]-piam*y[a-1,18,i]-pisd*y[a-1,18,i]
		y[a,19,i] <- y[a-1,19,i]+piam*y[a-1,18,i]-pimd*y[a-1,19,i]-pisd*y[a-1,19,i]
		y[a,20,i] <- y[a-1,20,i]+pisd*nmalH[a-1,i]+pimd*y[a-1,20,i]

		#calculation of sample size according to risk group and sex for year a of follow-up; only those alive are considered excluding those in the state of death (with 		indices 5, 10, 15, 20)
		
		nfemL[a,i] <- y[a,1,i]+y[a,2,i]+y[a,3,i]+y[a,4,i]
		nfemH[a,i] <- y[a,6,i]+y[a,7,i]+y[a,8,i]+y[a,9,i]
		nmalL[a,i] <- y[a,11,i]+y[a,12,i]+y[a,13,i]+y[a,14,i]
		nmalH[a,i] <- y[a,16,i]+y[a,17,i]+y[a,18,i]+y[a,19,i]

		#the probabilities of selecting a male/female partner from the high- or low-risk group, respectively, in year a of follow-up
		gmalH[a,i] <- (rmalH*nmalH[a,i])/(rmalH*nmalH[a,i]+rmalL*nmalL[a,i])
		gmalL[a,i] <- 1-gmalH[a,i]
		gfemH[a,i] <- (rfemH*nfemH[a,i])/(rfemH*nfemH[a,i]+rfemL*nfemL[a,i])
		gfemL[a,i] <- 1-gfemH[a,i]

		#weighted average of STI prevalence in males in year a of follow-up 
		pmal[a,i] <- gmalH[a,i]*(y[a,17,i]/nmalH[a,i])+gmalL[a,i]*(y[a,12,i]/nmalL[a,i])
		#weighted average of STI prevalence in females in year a of follow-up 
		pfem[a,i] <- gfemH[a,i]*(y[a,7,i]/nfemH[a,i])+gfemL[a,i]*(y[a,2,i]/nfemL[a,i])
		#force of infection: function of STI transmission probability beta, partner acquisition rates in low-risk females rfemL (accordingly for the other sex and risk group), and sex-specific population prevalence pmal and pfem, in year a of follow-up)

		lambdafemL[a,i] <-1-exp(-(pi2[i]*(1-eff2[i])*beta*rfemL*pmal[a,i]+(1-pi2[i])*beta
				*rfemL*pmal[a,i]))
		lambdafemH[a,i] <-1-exp(-(pi2[i]*(1-eff2[i])*beta*rfemH*pmal[a,i]+(1-pi2[i])*beta
				*rfemH*pmal[a,i]))
		lambdamalL[a,i] <-1-exp(-(pi2[i]*(1-eff2[i])*beta*rmalL*pfem[a,i]+(1-pi2[i])*beta
				*rmalL*pfem[a,i]))
		lambdamalH[a,i] <-1-exp(-(pi2[i]*(1-eff2[i])*beta*rmalH*pfem[a,i]+(1-pi2[i])*beta
				*rmalH*pfem[a,i]))
	}
}

#poisson-gamma model to update informative priors on partner acquisition rates by data
#datfemH: number of yearly partners in high-risk females, accordingly for the other sex and risk group
#rfemH: partner acquisition rate in high-risk females, accordingly for the other sex and risk group
#popsize: population size of the simulated data on yearly partners; we assume that 500 individuals were interviewed on their yearly number of partners

for(i in 1:popsize){
	datfemH[i] ~ dpois(rfemH)
	datfemL[i] ~ dpois(rfemL)
	datmalH[i] ~ dpois(rmalH)
	datmalL[i] ~ dpois(rmalL)
}

#informative priors on partner acquisition rates
rfemH ~ dgamma(0.1,0.1)
rfemL ~ dgamma(0.1,0.1)
rmalH ~ dgamma(0.1,0.1)
rmalL ~ dgamma(0.1,0.1)

#informative prior on reproduction rate chi.
chi~dgamma(1111.1,111111.1)

#informative prior on transition probability infected --> asymptomatic 
piia~dbeta(5119.2, 1279.8)

#informative prior on transition probability asymptomatic --> morbid 
piam~dbeta(1842.66,18631.34)

#informative prior on transition probability morbid --> dead
pimd~dbeta(1535.96,36863.04)

#informative prior on transition probability susceptible --> dead
pisd~dbeta(156.1714,312186.6)

#beta-binomial model on probability STI transmission. r.beta: number of events, n.beta: sample size
r.beta ~ dbin(beta,n.beta)
#minimally-informative prior on beta
beta ~ dbeta(0.5,0.5)

#beta-binomial model on probability of diagnosis. r.diag: number of events, n.diag: sample size
r.diag ~ dbin(diag,n.diag)
#minimally-informative prior on diag
diag ~ dbeta(0.5,0.5)

#beta-binomial model on probability of screening. r.screen: number of events, n.screen: sample size
r.screen ~ dbin(screen,n.screen)
#minimally-informative prior on screen
screen ~ dbeta(0.5,0.5)

#beta-binomial model on vaccine coverage. r.pi: number of events, n.pi: sample size
r.pi ~ dbin(pi,n.pi)
#minimally-informative prior on pi
pi ~ dbeta(0.5,0.5)

#beta-binomial model on vaccine efficacy. r.eff: number of events, n.eff: sample size
r.eff ~ dbin(eff,n.eff)
#minimally-informative prior on eff
eff ~ dbeta(0.5,0.5)

#informative priors on costs and utilities 
tau.screen<-1/pow(0.693,2)
c.screen ~ dlnorm(2.996,tau.screen)
tau.vac<-1/pow(0.07986607,2)
c.vac ~ dlnorm(5.518352,tau.vac)
tau.test.sti<-1/pow(0.03,2)
c.test.sti ~ dlnorm(2.996,tau.test.sti)
tau.test.blood<-1/pow(0.03,2)
c.test.blood ~ dlnorm(3.401,tau.test.blood)
tau.trt<-1/pow(0.8325546,2)
c.trt ~ dlnorm(4.258597,tau.trt)
tau.trt.dis<-1/pow(0.1980422,2)
c.trt.dis ~ dlnorm(6.194998,tau.trt.dis)
tau.gp<-1/pow(0.02,2)
c.gp ~ dlnorm(3.912,tau.gp)

#utility of infected
u.inf ~ dbeta(1469.3,629.7)
#utility of asymptomatic
u.asym ~ dbeta(1439.4,959.6)
#utility of morbid
u.morb ~ dbeta(629.7,1469.3)}
\end{lstlisting}

\subsection*{\texttt{WinBUGS} model}
\begin{lstlisting}
###WinBUGS model
model {
#initialization
female[1,1,1,1] <- 399760; female[1,1,1,2] <- 240; female[1,1,1,3] <- 0; 
female[1,1,1,4] <- 0; female[1,1,1,5] <- 0; female[1,2,1,1] <- 99940; 
female[1,2,1,2] <- 60; female[1,2,1,3] <- 0; female[1,2,1,4] <- 0; 
female[1,2,1,5] <- 0;

male[1,1,1,1] <- 399760; male[1,1,1,2] <- 240; male[1,1,1,3] <- 0; 
male[1,1,1,4] <- 0; male[1,1,1,5] <- 0; male[1,2,1,1] <- 99940; 
male[1,2,1,2] <- 60; male[1,2,1,3] <- 0; male[1,2,1,4] <- 0; 
male[1,2,1,5] <- 0;

female[1,1,2,1] <- 399760; female[1,1,2,2] <- 240; female[1,1,2,3] <- 0; 
female[1,1,2,4] <- 0; female[1,1,2,5] <- 0; female[1,2,2,1] <- 99940; 
female[1,2,2,2] <- 60; female[1,2,2,3] <- 0; female[1,2,2,4] <- 0; 
female[1,2,2,5] <- 0;

male[1,1,2,1] <- 399760; male[1,1,2,2] <- 240; male[1,1,2,3] <- 0; 
male[1,1,2,4] <- 0; male[1,1,2,5] <- 0; male[1,2,2,1] <- 99940; 
male[1,2,2,2] <- 60; male[1,2,2,3] <- 0; male[1,2,2,4] <- 0; 
male[1,2,2,5] <- 0;

#calculation of sample size according to risk-group and sex for year 1 of follow-up; only those alive are considered
#excluding those in the state of death with indices 5, 10, 15, 20)
nfem[1,1,1] <- female[1,1,1,1]+female[1,1,1,2]+female[1,1,1,3]+female[1,1,1,4]
nmal[1,1,1] <- male[1,1,1,1]+male[1,1,1,2]+male[1,1,1,3]+male[1,1,1,4]
nfem[1,2,1] <- female[1,2,1,1]+female[1,2,1,2]+female[1,2,1,3]+female[1,2,1,4]
nmal[1,2,1] <- male[1,2,1,1]+male[1,2,1,2]+male[1,2,1,3]+male[1,2,1,4]

nfem[1,1,2]<-nfem[1,1,1]
nmal[1,1,2]<-nmal[1,1,1]
nfem[1,2,2]<-nfem[1,2,1]
nmal[1,2,2]<-nmal[1,2,1]

#the probabilities of selecting a male/female partner from the high- or low-risk group, respectively, in year 1 of follow-up
gmalH[1,1] <- (rmalH*nmal[1,2,1])/(rmalH*nmal[1,2,1]+rmalL*nmal[1,1,1])
gmalL[1,1] <- 1-gmalH[1,1]
gfemH[1,1] <- (rfemH*nfem[1,2,1])/(rfemH*nfem[1,2,1]+rfemL*nfem[1,1,1])
gfemL[1,1] <- 1-gfemH[1,1]

gmalH[1,2]<-gmalH[1,1]
gmalL[1,2]<-gmalL[1,1]
gfemH[1,2]<-gfemH[1,1]
gfemL[1,2]<-gfemL[1,1]

#calibration
dat.sus.femH[1] ~ dpois(female[1,2,1,1])
dat.inf.femH[1] ~ dpois(female[1,2,1,2])
dat.asy.femH[1] ~ dpois(female[1,2,1,3])
dat.mor.femH[1] ~ dpois(female[1,2,1,4])
    
dat.sus.malH[1] ~ dpois(male[1,2,1,1])
dat.inf.malH[1] ~ dpois(male[1,2,1,2])
dat.asy.malH[1] ~ dpois(male[1,2,1,3])
dat.mor.malH[1] ~ dpois(male[1,2,1,4])

#dat.sus.femH: simulated data on susceptible high-risk females, used for calibration in year 1. The parameter female[n,2,1,1] contains the number of susceptible high-risk females in year 1 and is calibrated by the data for year 1. The approach for high-risk females in the states infected, asymptomatic and morbid as well as for high-risk males in the same states is conducted accordingly.

for (n in 2:5){
	conssusfemH[n] <- max(0.1,female[n,2,1,1])
	dat.sus.femH[n] ~ dpois(conssusfemH[n])
    
	consinffemH[n] <- max(0.1,female[n,2,1,2])
	dat.inf.femH[n] ~ dpois(consinffemH[n])
    
	consasyfemH[n] <- max(0.1,female[n,2,1,3])
	dat.asy.femH[n] ~ dpois(consasyfemH[n])
    
	consmorfemH[n] <- max(0.1,female[n,2,1,4])
	dat.mor.femH[n] ~ dpois(consmorfemH[n])
    
	conssusmalH[n] <- max(0.1,male[n,2,1,1])
	dat.sus.malH[n] ~ dpois(conssusmalH[n])
    
	consinfmalH[n] <- max(0.1,male[n,2,1,2])
	dat.inf.malH[n] ~ dpois(consinfmalH[n])
    
	consasymalH[n] <- max(0.1,male[n,2,1,3])
	dat.asy.malH[n] ~ dpois(consasymalH[n])
    
	consmormalH[n] <- max(0.1,male[n,2,1,4])
	dat.mor.malH[n] ~ dpois(consmormalH[n])
}

#state allocation algorithm for years 2-5 of follow-up
#year a
for (a in 1:A){
	#sexual behaviour group g (1=low-risk, 2=high-risk)
	for (g in 1:2){
	#intervention i (1=screening-only, 2=vaccination)
		for (i in 1:2){

			#transition probability matrix. Lambda: transition probabilities in females, Kappa: transition probabilties in males
			Lambda[a,g,i,1,3]<-0
			Lambda[a,g,i,1,4]<-0
			Lambda[a,g,i,1,5]<-pisd

			Lambda[a,g,i,2,1]<-0
			Lambda[a,g,i,2,2]<-1-piia-pisd
			Lambda[a,g,i,2,3]<-piia
			Lambda[a,g,i,2,4]<-0
			Lambda[a,g,i,2,5]<-pisd

			Lambda[a,g,i,3,1]<-0
			Lambda[a,g,i,3,2]<-0
			Lambda[a,g,i,3,3]<-1-piam-pisd
			Lambda[a,g,i,3,4]<-piam
			Lambda[a,g,i,3,5]<-pisd

			Lambda[a,g,i,4,1]<-0
			Lambda[a,g,i,4,2]<-0
			Lambda[a,g,i,4,3]<-0
			Lambda[a,g,i,4,4]<-1-pisd-pimd
			Lambda[a,g,i,4,5]<-pisd+pimd

			Lambda[a,g,i,5,1]<-0
			Lambda[a,g,i,5,2]<-0
			Lambda[a,g,i,5,3]<-0
			Lambda[a,g,i,5,4]<-0
			Lambda[a,g,i,5,5]<-1

			Kappa[a,g,i,1,3]<-0
			Kappa[a,g,i,1,4]<-0
			Kappa[a,g,i,1,5]<-pisd

			Kappa[a,g,i,2,1]<-0
			Kappa[a,g,i,2,2]<-1-piia-pisd
			Kappa[a,g,i,2,3]<-piia
			Kappa[a,g,i,2,4]<-0
			Kappa[a,g,i,2,5]<-pisd

			Kappa[a,g,i,3,1]<-0
			Kappa[a,g,i,3,2]<-0
			Kappa[a,g,i,3,3]<-1-piam-pisd
			Kappa[a,g,i,3,4]<-piam
			Kappa[a,g,i,3,5]<-pisd

			Kappa[a,g,i,4,1]<-0
			Kappa[a,g,i,4,2]<-0
			Kappa[a,g,i,4,3]<-0
			Kappa[a,g,i,4,4]<-1-pisd-pimd
			Kappa[a,g,i,4,5]<-pisd+pimd

			Kappa[a,g,i,5,1]<-0
			Kappa[a,g,i,5,2]<-0
			Kappa[a,g,i,5,3]<-0
			Kappa[a,g,i,5,4]<-0
			Kappa[a,g,i,5,5]<-1
		}
	}
}

for (i in 1:2){
	#vaccine coverage and efficacy per intervention i
	pi2[i]<-pi*step(i-1.1)+(1-step(i-1.1))
  eff2[i]<-eff*step(i-1.1)

	for (a in 2:A){
		#weighted average of STI prevalence in males and females in year a of follow-up 	
		pmal[a,i] <- gmalH[a-1,i]*(male[a-1,2,i,2]/nmal[a-1,2,i])+gmalL[a-1,i]
				*(male[a-1,1,i,2]/nmal[a-1,1,i])
    pfem[a,i] <- gfemH[a-1,i]*(female[a-1,2,i,2]/nfem[a-1,2,i])+gfemL[a-1,i]
    		*(female[a-1,1,i,2]/nfem[a-1,1,i])
	
		#force of infection and probability of remaining susceptible
	  Lambda[a,1,i,1,2]<-1-exp(-(pi2[i]*(1-eff2[i])*beta*rfemL*pmal[a,i]+(1-pi2[i])*beta
	  		*rfemL*pmal[a,i]))
	  Lambda[a,1,i,1,1]<-1-Lambda[a,1,i,1,2]-pisd
	  		
		Lambda[a,2,i,1,2]<-1-exp(-(pi2[i]*(1-eff2[i])*beta*rfemH*pmal[a,i]+(1-pi2[i])*beta
				*rfemH*pmal[a,i]))
		Lambda[a,2,i,1,1]<-1-Lambda[a,2,i,1,2]-Lambda[a,2,i,1,5]
			
		Kappa[a,1,i,1,2]<-1-exp(-(pi2[i]*(1-eff2[i])*beta*rmalL*pfem[a,i]+(1-pi2[i])*beta
				*rmalL*pfem[a,i]))
	  Kappa[a,1,i,1,1]<-1-Kappa[a,1,i,1,2]-Kappa[a,1,i,1,5]
			
		Kappa[a,2,i,1,2]<-1-exp(-(pi2[i]*(1-eff2[i])*beta*rmalH*pfem[a,i]+(1-pi2[i])*beta
				*rmalH*pfem[a,i]))
		Kappa[a,2,i,1,1]<-1-Kappa[a,2,i,1,2]-Kappa[a,2,i,1,5]
		
    #state allocation algorithm
    for (g in 1:2){
    	#accounting for proliferation in population
    	female[a,g,i,1] <-(inprod(female[a-1,g,i,], Lambda[a,g,i,,1]))+chi*nfem[a-1,g,i]
			male[a,g,i,1] <-(inprod(male[a-1,g,i,], Kappa[a,g,i,,1]))+chi*nmal[a-1,g,i]
    
    	#inner product of number of people in states and transition probability matrices
    	for (s in 2:5){
				female[a,g,i,s] <-inprod(female[a-1,g,i,], Lambda[a,g,i,,s])
				male[a,g,i,s] <-inprod(male[a-1,g,i,], Kappa[a,g,i,,s])
    	}
    
    	#calculating sample size of females and males
    	nfem[a,g,i] <- female[a,g,i,1]+female[a,g,i,2]+female[a,g,i,3]+female[a,g,i,4]
			nmal[a,g,i] <- male[a,g,i,1]+male[a,g,i,2]+male[a,g,i,3]+male[a,g,i,4]
		}
		
		#calculating probabilities of selecting partners from the two risk groups
		gmalH[a,i] <- (rmalH*nmal[a,2,i])/(rmalH*nmal[a,2,i]+rmalL*nmal[a,1,i])
    gmalL[a,i] <- 1-gmalH[a,i]
    gfemH[a,i] <- (rfemH*nfem[a,2,i])/(rfemH*nfem[a,2,i]+rfemL*nfem[a,1,i])
    gfemL[a,i] <- 1-gfemH[a,i]
	}
}
			
#poisson-gamma model to update informative priors on partner acquisition rates by data
#datfemH: number of yearly partners in high-risk females, accordingly for the other sex and risk-group
#rfemH: partner acquisition rate in high-risk females, accordingly for the other sex and risk-group
#popsize: population size of the simulated data on yearly partners; we assume that 500 individuals were interviewed on their yearly number of partners

for(i in 1:popsize){
	datfemH[i] ~ dpois(rfemH)
	datfemL[i] ~ dpois(rfemL)
	datmalH[i] ~ dpois(rmalH)
	datmalL[i] ~ dpois(rmalL)
}

#informative priors on partner acquisition rates
	rfemH ~ dgamma(0.1,0.1)
	rfemL ~ dgamma(0.1,0.1)
	rmalH ~ dgamma(0.1,0.1)
	rmalL ~ dgamma(0.1,0.1)

#informative prior on reproduction rate chi
chi~dgamma(1111.1,111111.1)

#informative prior on transition probability infected --> asymptomatic
piia~dbeta(5119.2, 1279.8)

#informative prior on transition probability asymptomatic --> morbid
piam~dbeta(1842.66,18631.34)

#informative prior on transition probability morbid --> dead
pimd~dbeta(1535.96,36863.04)

#informative prior on transition probability susceptible --> dead
pisd~dbeta(156.1714,312186.6)

#beta-binomial model on probability STI transmission. r.beta: number of events, n.beta: sample size
r.beta ~ dbin(beta,n.beta)
#minimally-informative prior on beta
beta ~ dbeta(0.5,0.5)
#beta ~dbeta(764.85,4334.15)

#beta-binomial model on probability of diagnosis. r.diag: number of events, n.diag: sample size
r.diag ~ dbin(diag,n.diag)
#minimally-informative prior on diag
diag ~ dbeta(0.5,0.5)

#beta-binomial model on probability of screening. r.screen: number of events, n.screen: sample size
r.screen ~ dbin(screen,n.screen)
#minimally-informative prior on screen
screen ~ dbeta(0.5,0.5)

#beta - binomial model on vaccine coverage
r.pi ~ dbin(pi,n.pi)
pi ~ dbeta(0.5,0.5)

#beta - binomial model on vaccine efficacy
r.eff ~ dbin(eff,n.eff)
eff ~ dbeta(0.5,0.5)

#costs and utilities
tau.screen<-1/pow(0.693,2)
c.screen ~ dlnorm(2.996,tau.screen)
tau.vac<-1/pow(0.07986607,2)
c.vac ~ dlnorm(5.518352,tau.vac)
tau.test.sti<-1/pow(0.03,2)
c.test.sti ~ dlnorm(2.996,tau.test.sti)
tau.test.blood<-1/pow(0.03,2)
c.test.blood ~ dlnorm(3.401,tau.test.blood)
tau.trt<-1/pow(0.8325546,2)
c.trt ~ dlnorm(4.258597,tau.trt)
tau.trt.dis<-1/pow(0.1980422,2)
c.trt.dis ~ dlnorm(6.194998,tau.trt.dis)
tau.gp<-1/pow(0.02,2)
c.gp ~ dlnorm(3.912,tau.gp)

#utility of infected
u.inf ~ dbeta(1469.3,629.7)
#utility of asymptomatic
u.asym ~ dbeta(1439.4,959.6)
#utility of morbid
u.morb ~ dbeta(629.7,1469.3)
}

\end{lstlisting}

\end{appendix}

\end{document}